\documentstyle[epsf,eqsecnum,aps,twocolumn]{revtex}

\begin{document}
\draft
\preprint{KMC--MD}
\title{Simulations of energetic beam deposition: from picoseconds to seconds}

\author{Joachim Jacobsen$^\dagger$, B.H. Cooper, and James P. Sethna}

\address{Laboratory of Atomic and Solid State Physics and Cornell Center
for Materials Research, Cornell University, Ithaca, NY 14853-2501, USA\break
\bigskip
\noindent
$^\dagger$Present address:
Haldor Topsoe A/S, Nymollevej 55, DK-2800 Lyngby, Denmark.
}

\date{\today}

\maketitle

\begin{abstract}

We present a new method for simulating crystal growth by energetic 
beam deposition. The method combines a Kinetic Monte--Carlo
simulation for the thermal surface diffusion with a small scale
molecular dynamics simulation of every single deposition event.
We have implemented the method using the effective medium theory
as a model potential for the atomic interactions, and present 
simulations for Ag/Ag(111) and Pt/Pt(111) for incoming energies up
to 35 eV. The method is capable of following the growth of several 
monolayers at realistic growth rates of 1 monolayer per second,
correctly accounting for both energy--induced atomic mobility 
and thermal surface diffusion. We find that the energy influences island
and step densities and can induce layer--by--layer growth. We find an
 optimal energy for layer--by--layer growth (25 eV for Ag), which
correlates with where the net impact-induced downward interlayer transport
is at a maximum.  A high step density is needed for energy induced
layer--by--layer growth, hence the effect
dies away at increased temperatures, where thermal surface diffusion 
reduces the step density. As part of the development of the method, we
present molecular dynamics simulations of single atom--surface collisions 
on flat
parts of the surface and near straight steps, we identify microscopic
mechanisms by which the energy influences the growth, and we discuss the nature
of the energy--induced atomic mobility.
\end{abstract}
\pacs{PACS numbers: 
% Ion and electron beam-assisted deposition; ion plating
81.15.Jj, 
% Thin film structure and morphology
68.55.-a, 
% Kinetics of defect formation and annealing
61.72.Cc}

\narrowtext 

\section{Introduction}

The overall goal of being able to manufacture nanoscale devices while
controlling chemical composition and crystal structure has been the
driving force behind a tremendous effort in numerous research
groups over the past decades, and the ability to grow crystal
surfaces in a 
layer-by-layer fashion has been a key issue~\cite{barabasi}.
The use of energetic particles has offered a promising possibility of gaining 
control of the growth process~\cite{smidt,greene}.

There are numerous examples where energetic beams have been used to
control and improve the properties of the grown materials. 
Ion Beam Assisted Deposition ~\cite{shindo}, 
Ion Beam Sputter Deposition~\cite{lee}
and Ion Beam Direct Deposition~\cite{karr,rabalais}, 
have been used to lower the epitaxial growth temperature and improve smoothness.
Increased control of interfacial roughness has been achieved through 
sputter deposition~\cite{fullerton93}, and Giant Magnetoresistance
has been improved both using both sputter
deposition~\cite{fullerton92,ueda} and using
direct ion beam deposition~\cite{nagamachi}.

Not surprisingly, there can be an optimal window for the energy per 
incoming particle: a certain energy is needed to increase atomic mobility
at the surface, but too high an energy can cause a drastic increase in
defect formation.
There is a growing literature on experiments which map out parameter
space to find optimal values for various systems, and the optimal
energy window is often found at relatively low energies.  For example,
Rabalais {\it et al.} found that for silicon ion--beam epitaxy very smooth 
growth was obtained using 20 eV particles~\cite{rabalais}.

There is an obvious need for improving our understanding of how the
energy influences the growth. What are the microscopic mechanisms by
which the energy changes the growth morphology? Recent experiments
have illustrated how the energetic ion--surface collisions can
influence submonolayer island densities during growth of a Pt(111)
surface by sputter deposition~\cite{kallf,michely} or Ion Beam
assisted deposition~\cite{esch}. Several experimental groups are
operating~\cite{karr,rabalais} or building equipment for Ion Beam
Direct Deposition 
with high control of beam energy and angle of
incidence~\cite{beamlines}, and we can
expect much more detailed experimental information on the effect of
the energy in the near future.

Molecular dynamics (MD) simulations using model potentials,
from which very detailed information on microscopic mechanisms can be obtained,
have proven 
to be a very useful tool in studying energetic
ion--surface~\cite{villarba,kitabatake,spraque,gilmore,kelchner,ohashi} and
cluster--surface~\cite{clustermd} collisions.
There are several questions that can be directly addressed using MD.
What kind of atomic
rearrangements occur in the ion--surface collisions as a function of
energy? Is ballistic motion or local heating the right picture of the
energy--induced mobility? What are the time
and length scales of the induced mobility? 
For example, Villarba and Jonsson~\cite{villarba} studied low energy 
(10 and 20 eV) impact of Pt atoms on a Pt(111) surface, and  identified
push--out events where atoms impinging close to descending steps are 
incorporated into the growing layer. The net effect is that the step edge
grows horizontally favoring a layer-by-layer growth mode. They found
that the non-thermal effect of the collision was over in a few
picoseconds, and the ranges of the collision--induced atomic rearrangements
were a few lattice sites.

One issue, which MD fails to address, and which is crucial
in understanding most experiments using energetic particles, is what
is the relative importance of collision--induced atomic mobility on the
one hand, and the thermal surface diffusion on the other?
MD simulations attempting to simulate the entire growth process must
use deposition rates on the order of $10^{10}$
monolayers per sec (ML/s)~\cite{spraque,gilmore,kelchner},
and neglect the effect
thermal surface diffusion on longer time--scales. Furthermore,
subsequent collision events might affect each other directly in ways 
they would not at the much lower experimental deposition rates (typically
0.01 - 1 ML/s). If we want to address this question of the relative
importance of collision--induced 
mobility and the thermal surface diffusion, we are faced with
a tremendous problem of time--scales. Each atom--surface collision is a
pico--second event, to resolve which you need femto--second numerical time
steps. At the other extreme, in the typical experimental situation there is
one such collision per lattice site per 1 to 100 seconds. 

In this paper we present results of a new simulation method capable of 
overcoming this 15 orders of magnitude gap in time--scales.
We have combined two well known simulation techniques in a
new way. We do a traditional Kinetic Monte--Carlo (KMC) simulation of the
thermal surface diffusion in between the rare collisions, and for
each of these collisions we do a small length--scale, short
time--scale molecular dynamics simulation. We call this a hybrid 
Kinetic Monte--Carlo Molecular Dynamics (KMC--MD) simulation. 
We will show the results of depositing several monolayers at 1ML/sec, while
treating every collision explicitly in MD simulations. As model systems,
we have chosen Ag(111) and Pt(111) homoepitaxy, and we use Effective
Medium Theory (EMT)~\cite{emt} as a model potential.
An idea similar to KMC--MD
has been used to simulate damage production in ion implantation of
silicon~\cite{rubia}.

We use effective medium theory, which is known to give a good qualitative
and to some extent quantitative description of these 
metals~\cite{emt,ptisl}.
We do not expect these simulations to accurately predict all features of
the growth by energetic beam deposition of Ag(111) and
Pt(111)~\cite{ptisl}.
However, we do hope to gain useful insight into
how the use of energetic particles can influence growth, 
microscopic mechanisms, energy--induced atomic mobility, and the
interplay between the energy--induced mobility and thermal surface diffusion.

The paper is organized as follows: In section~\ref{sec:method}, we
describe the KMC--MD method in detail, going through how the KMC part
of the simulation is done (section~\ref{sec:methodKMC}), how the 
MD part is done (section~\ref{sec:methodmd}), how to go from the
discrete KMC  to the continuous MD simulation
(section~\ref{sec:methodKMCtomd}), and back (section~\ref{sec:methodmdtoKMC}).
The results section, part~\ref{sec:results}, is divided into two main parts
-- MD simulations of single energetic atom--surface collisions
(section~\ref{sec:resultsmd}) and KMC--MD simulations of the
entire growth process (section~\ref{sec:resultskmcmd}). In
section~\ref{sec:resultsmd} we start our discussion of the choice of
MD system size (section~\ref{sec:resultsmdsize}), where we show that large MD
systems and Langevin damping on the boundaries are essential to avoid
unphysical reflections of the supersonic shock-wave induced by the impact.
We then go through
some important energy induced microscopic mechanisms we find for
Ag$\rightarrow$Ag(111) and Pt$\rightarrow$Pt(111) energetic impacts,
focusing on the energy-induced upward and downward interlayer mobility at
straight step edges. We round up giving a short summary and discussion of
these mechanisms in
section~\ref{sec:resultsmdsum}. In section~\ref{sec:resultskmcmd}, we
present our KMC--MD results for the submonolayer structure and for the
surface roughness after the deposition of a few atomic layers; first
for Ag/Ag(111) and then for Pt/Pt(111). We find the smoothest growth
when both the net downward interlayer mobility and the step edge density
are large.  In section~\ref{sec:discres}
we give a concluding discussion of our results for simulations of
growth by energetic beam deposition, and in
section~\ref{sec:discmeth} we give a general discussion of the KMC--MD
method. 

\section{The KMC--MD Method}
\label{sec:method}

In the introduction above, we stated that a main difficulty in 
simulating crystal growth by energetic depositions is the tremendous
gap in time scales between the picosecond collisions and the
deposition, typically slower than one impact per lattice site per
second. The idea behind the KMC--MD method is precisely that the
depositions are rare events. For most of the time during the crystal
growth, there is no non--thermal atomic mobility on the surface. Only
the thermal diffusion of atoms is active. Kinetic Monte--Carlo
is a very efficient way of evolving the surface for this time in
between the energetic collisions, given a model of the thermal
diffusion. 

Once in a great while (on the time scale of diffusion) an energetic collision
occurs. These collisions may displace the incoming atom
from the impact site, as well as rearrange the surface atoms. But for
homoepitaxial metal systems, the energy transfer is very efficient,
and the incoming atom generally slows down very quickly. Also, 
in the crystalline environment, the excess energy dissipates away very
quickly due to phonons. Consequently, non-thermal mobility is limited
to a very short time, $\tau_E$, after the collision. Furthermore, for
not too high incoming energies, the collision--induced mobility is
fairly short ranged. This makes MD of the collisions a
suitable method. 
 
In the KMC--MD method, we evolve time using a KMC simulation in a
standard way~\cite{kmcgeneral,kmcpt2d3d,kmcexample,kmcagpt}.
Diffusion processes happen sequentially according to
their relative rates and, with a probability proportional to the
deposition rate, new atoms are introduced above the surface with a
specified kinetic energy and angle and aimed at a random point on the
surface. When this happens, we set up an MD simulation to correctly
describe the atom--surface collision. We include only the local region
in the vicinity of the impact site in the MD simulation, and run for
only a short time. We then feed the end result back into the KMC
simulation, and continue. In sections~\ref{sec:method} A-D below,
we describe the details of the different parts of the KMC--MD method.

We model the atomic interactions using effective medium
theory~\cite{emt} --- in the MD simulations of the energetic collisions 
we use the EMT forces, and in the KMC--simulations we use a comprehensive
set of EMT energy barriers for various atomic diffusion processes. 

\subsection{The KMC lattice simulation}
\label{sec:methodKMC}

Kinetic Monte--Carlo has become a standard method
for doing lattice simulations of crystal 
growth~\cite{kmcgeneral,kmcpt2d3d,kmcexample,kmcagpt}. 
In short, we make a complete table of the active atomic diffusion processes
at the given temperature. For each type of diffusion process we
evaluate its rate as $r_i = \nu \exp{(-E_i/k_B T)}$, where we assume
a common prefactor of $\nu = 10^{12} s^{-1}$ for simplicity.
Table~\ref{tab:modbar} lists the processes we include together with
their EMT energy barriers $E_i$ for the two systems (Ag/Ag(111) and
Pt/Pt(111)). The surface atomic configuration is specified
by the occupancy by atoms of the fcc(111) lattice sites.
For every atom 
on the surface we examine if it can potentially make a lattice jump
in any of the processes from table~\ref{tab:modbar}, and if so we 
add these specific atomic jumps, and their rate $r_i$ to a list of
all the possible potential diffusion processes the given surface 
configuration can evolve by. Also included in the list is the deposition
of a new atom --- with a rate given by the deposition flux times the surface
area.
 Now, in every loop of the program,
with a probability proportional to its rate, one particular atomic 
jump is chosen from the list of potential processes. The surface configuration
and the list of potential processes is then updated accordingly, taking
advantage of the fact that there will be only local changes.
Because there is a lattice jump or a new deposition in every loop
of the program, KMC is a very efficient method, and 
can bridge one gap of time--scales: the gap between fast thermal diffusion
processes ($10^{6}$ to $10^{9}$ per sec.) and the deposition rate. 
We depart from the standard way of doing crystal growth with KMC in the
way we handle deposition events.
Usually in KMC simulations when a deposition event is chosen to happen, 
a new atom is introduced at a random surface lattice site.
Instead we do something new --- a complete MD simulation of the
atom--surface collision with the specified energy and angle and an
impact parameter for the incoming atom chosen at random.

In simulations of thermal crystal growth and relaxation the KMC method
has been used in two ways: the energy barriers giving the rates can be 
obtained from a model potential energy, or they can be used as 
adjustable parameters to fit growth experiments or to study the effect of 
a particular process on
the resulting surface morphology. In our case, to make the two parts of the
simulation consistent with one another, the KMC energy barriers must be
obtained from the potential energy used in the MD simulations.

\subsection{From discrete KMC to continuous MD}
\label{sec:methodKMCtomd}

When the KMC algorithm chooses a deposition event, a random point in a
plane above the surface is picked. From this point, a straight line at
the ion beam angle is then followed until it crosses a horizontal
(111)--plane near (within a surface unit cell of) a lattice site
occupied by an atom, which we will call the impact atom.
With respect to this expected impact atom,
the cluster is set up based on the occupancy of
sites in the KMC--simulation, as discussed below. 

\begin{table}[htbp]
  \begin{center}
    \begin{tabular}{lccrrr}
 {\em Metal } & & & & Ag & Pt \\
      {\em Terrace diffusion}  & & & & &     \\ 
      \hspace*{5mm}  Diffusion of monomers & & & & 67&158             \\ 
      \hspace*{5mm}  Diffusion away from descending steps & & & &---&208   \\ 
      \hspace*{5mm}  Diffusion of dimers  & & & & 132&220             \\ 
      \hspace*{5mm}  Diffusion of vacancies & & & & 540&690             \\ 
      {\em Detachment} & & & &                 \\ 
      \hspace*{5mm}  Dissociation from 1 neighbor  & & & & 315&500 \\
      {\em Edge diffusion}    &$N_i$ &$T\! S$ &$N_f$& \\ 
      \hspace*{5mm}Corner diffusion              &1 & A &$\geq$ 1 & 77&220 \\
      \hspace*{5mm}Corner diffusion              &1 & B &$\geq$ 1 & 132&130 \\
      \hspace*{5mm}Step to corner                &2 & A &       1 & 257&510  \\ 
      \hspace*{5mm}Step to corner                &2 & B &       1 & 317&410 \\ 
      \hspace*{5mm}Step Diffusion                &2 & A &$>$    1 & 221&450 \\
      \hspace*{5mm}Step Diffusion                &2 & B &$>$    1 & 296&390 \\ 
      \hspace*{5mm}Kink to corner                &3 & A &       1 & 423&740 \\ 
      \hspace*{5mm}Kink to corner                &3 & B &       1 & 478&650 \\ 
      \hspace*{5mm}Kink to step                  &3 & A &$>$    1 & 387&680 \\ 
      \hspace*{5mm}Kink to step                  &3 & B &$>$    1 & 457&630 \\ 
      {\em Inter--layer diffusion}  & & & &        &                    \\ 
      \hspace*{5mm}Descent at straight step & & & &  240&400             \\ 
      \hspace*{5mm}Descent next to kink at B-step & & & & ---&270  \\      
    \end{tabular}
  \end{center}
  \caption{Effective Medium energy barriers in meV used in the KMC
      simulations. In detachment processes
      an atom stays at the surface, but dissociates 
      from in--layer neighbors. Edge diffusion is for atoms moving along
      island edges.
      $N_i$ and $N_f$ are the initial and final in--layer
      coordinations of the moving
      atom. The transition state is labeled $A$ or $B$
      if the motion is along a
      (100) or (111) micro-facet.
      Dimers can diffuse via a single atom mechanism in 
      which one atom moves along the edge of the other.
      The atomic moves are illustrated in detail in
      refs.~\protect\cite{kmcptisl}.
      The ``---'' for Ag
      indicates that the barrier is not distinguished from the one 
      immediately above.}
  \label{tab:modbar}
\end{table}

Before introducing the energetic
atom, the system is equilibrated at the specified temperature. We
find that an efficient way to do this is to give every atom a velocity
picked from a Maxwell--Boltzmann distribution corresponding to twice
the temperature. Since the atoms start out on perfect lattice
sites, this gets the total energy per atom roughly right. We then need
only to run for 0.27 ps to get a reasonably equilibrated system. 
Finally, the atom is placed three horizontal (111) planes upward
along the straight line introduced above, and is started with
the specified kinetic energy.

\subsection{The MD continuous simulations}
\label{sec:methodmd}

\begin{figure}[htbp]
  \begin{center}
    \leavevmode
    \epsfxsize=8cm
    \epsffile{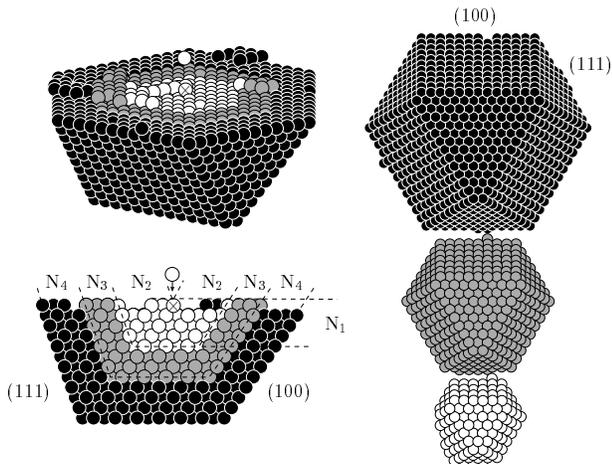}
    \caption{Cluster setup used in the MD simulations. The MD--atoms
    (white) have zero Langevin friction coefficients, the Langevin atoms
    (gray) have positive Langevin friction coefficients, and the
    static  atoms (black) are immobile and fixed at lattice
    positions. Left: view 
    from glancing angle, and view 
    of a vertical cut. Right: view from below of the 3 types of atoms.
    We include atoms in shells bounded by
    (111) and (100) planes at specified distances to the impact atom
    (marked with $\times$). 
    Two integers specify the region of MD--atoms
    atoms: $\rm N_1$ is the depth in number of horizontal (111) planes
    and $\rm N_2$ is lateral size in number of (111) or (100) planes, as
    shown. The region of Langevin  atoms is the surrounding $\rm N_3$
    number of (111) or (100) planes, and likewise the region of static
    atoms is the surrounding $\rm N_4$ number of (111) or (100)
    planes. The shown setup has $\rm N_1=5$, $\rm N_2=6$, $\rm N_3=3$,
    and $\rm N_4=4$, or $\bar{\rm N}=(5,6,3,4)$. 
    ($\rm N_1=\rm N_2=1$ would give exactly 1 MD--atom).
}
    \label{fig:mdsetup}
  \end{center}
\end{figure}

Following DePristo and Metiu~\cite{depristo} we set up a cluster
consisting of three types of atoms. The atoms in the immediate
vicinity of the impact site evolve classically according to the EMT
forces only (we call these MD--atoms). Surrounding 
the MD--atoms we have a shell of Langevin atoms, which are
subject to the EMT forces, and in addition to a friction force and a
randomly fluctuating force. To fix the geometry a shell of static
atoms surrounds the shell of Langevin atoms
(Figure~\ref{fig:mdsetup}). The equations of motion for the dynamic atoms
are: 

\begin{equation}
m_i {d^2r_i \over dt^2} = - \nabla_{r_i} V_{EMT}(\{r\}) - 
m_i \xi_i {dr_i \over dt} + f_i(t)
\label{eq:lngvn}
\end{equation}
where $r_i$ and $m_i$ are the three Cartesian coordinates and the mass
of atom $i$,
$V_{EMT}(\{r\})$ is the EMT potential energy as a function of {\em all}
atomic coordinates $\{r\}$, $\xi_i$ is the Langevin friction
coefficient of the atom $i$ (the use of the index will be modified below), 
$t$ is time and $f(t)$ is the fluctuating
force obeying the fluctuation--dissipation theorem:

\begin{equation}
\langle f_i(t)f_i(0) \rangle = 2\xi_i m_i kT \delta (t).
\end{equation}
The central MD--atoms have zero friction coefficient, $\xi_i =0$. 
We integrate equation~\ref{eq:lngvn} in time using the algorithm
proposed by Allen and Tildesley~\cite{allen}, which reduces to the
Verlet algorithm for $\xi_i =0$. 

Having a shell of Langevin atoms which surrounds the
MD--atoms serves several purposes.
It allows us to equilibrate the system at a specified
temperature before shooting in the energetic atom; it mimics the
contact of the MD--atoms with an infinite heat bath (the crystal),
ensuring that the
deposited energy doesn't permanently heat up the system; and
furthermore, as we shall see later, it allows us to use a smaller
system size without unphysical reflections of energy from the
boundaries affecting the atomic motion near the impact site.
We will return to a discussion of this latter point in
section~\ref{sec:resultsmd}. We always use $N_3 = 3$ and $N_4 = 4$,
except for a test calculation with no Langevin atoms ($N_3 =
0$). Having 3 atomic planes of Langevin atoms, we use the possibility of
having different Langevin coefficients for each plane. We label these
coefficients $\bar\xi = (\xi_1, \xi_2, \xi_3)$, where $\xi_1$ is for the
plane neighboring the MD--atoms, and  $\xi_3$ is for the plane
neighboring the static atoms (fig.~\ref{fig:mdsetup}).

Apart from how the cluster boundaries affect the energy dissipation
away from the impact site, another criterion for choosing the size of a
cluster simulation is the range 
of the transient mobility induced by the collisions. 
In the energetic collisions, atoms are usually displaced a few
lattice sites from
the point of impact, and it is important that these atoms stay in
contact with the region of MD--atoms. Occasionally,
incoming atoms do not stick to the
surface, but are reflected back. For rare impact parameters on step
edges, incoming atoms may travel rather long distances
along the surface before sticking and thermalizing.
In long
simulations with thousands of depositions such events will be
encountered. However, at the end of an MD simulation where the
lattice sites of the atoms have to be identified, we check if
any dynamic atoms left the physical region of the cluster calculation.
Our KMC--MD system sizes were sufficiently large that 
this happened in fewer than 2\% of the deposition events.

Every MD simulation is followed for 5 ps. After this time, 
the average kinetic energy per atom in the local impact region is once
again $\sim 3/2\, k T$, and all further
atomic mobility will be thermally activated and well--described
by the KMC lattice simulation.

We find it important to use a variable time step in the
MD simulations of the collisions. However, based on
Figures~\ref{fig:eref_pop} and~\ref{fig:avekin},
we choose to do this in the simplest possible way.
The very high kinetic
energies are all found within the first 0.5 ps of the collisions, and
after this time the atoms have much more moderate velocities. We use
the Verlet algorithm, with a time step of $dt_1 = 1.08$ fs for the first 0.5
ps, and subsequently we use $dt_2 = 5.4$ fs for the remaining 4.5 ps. By
comparing to much shorter time steps, we have checked that this
scheme, for energies up to 30 eV, is sufficient to correctly identify
the lattice sites of all atoms after 5 ps, whereas using $dt_2$
for the entire collision would give inaccurate results.

\subsection{From continuous MD to discrete KMC}
\label{sec:methodmdtoKMC}

After having the done the MD simulation for the local vicinity of the
impact, when all atomic mobility is once again thermal, we need to
map the end result of the continuous simulation back to the lattice 
description of the KMC part of the simulation. In the thermal surface 
configuration every atom is oscillating around a local minimum of the
EMT potential energy surface. To identify the binding site of each atom,
we simply minimize the energy with respect to the atomic coordinates.
This uniquely identifies an fcc lattice site for almost all the atoms.

However, there is one exception to this. The presence
of hcp binding sites on the fcc(111) surface give rise to
off--lattice local energy minima which can trap atoms.
In EMT,
an isolated adatom at the surface has essentially the same energy
when at an fcc and an hcp site, and has roughly equal probabilities 
of occupying the two types of sites. As a result, during growth islands 
can nucleate on hcp sites as well as fcc sites, which would possibly make
the coalescence of each new layer a very complicated process involving
surface dislocations separating regions of hcp and fcc stacking sequences.
However, low temperature growth experiments for Ag/Ag(111) and Pt/Pt(111)
do not show the complicated behavior expected if islands of substantial
size had significant probabilities of being on hcp sites --- for
Ag/Ag(111) the growth is known to proceed on the fcc sites~\cite{hcpag}.
More accurate total energy calculations for
Pt/Pt(111)~\cite{hcptheory} finds a significantly increased energy for
the adatom at the hcp site,  suggesting
that the strong binding at hcp sites is an artifact of the
effective medium theory,
due to the lack of dependence of the energy on relative bond angles.
For these reasons, and because it would make the KMC simulation
infeasible, we do {\it not} allow atoms to occupy hcp sites. 

Since the hcp sites are unfavorable, and thermal diffusion will rapidly
shift atoms from hcp metastable sites to neighboring fcc sites,
we need a procedure which takes the atoms on hcp sites in the
energy minimized configuration, and puts them on one of the three 
neighboring fcc sites. 
When doing this the important thing is not
to break any bonds if, for example, a dimer on hcp sites has to be displaced.
For each atom that ended up on an hcp site after the energy minimization, 
we move it to the neighboring
vacant fcc site of highest in--layer coordination, or a random choice
among these sites if there are several equivalent ones.

With all atoms occupying fcc sites, we can continue the KMC simulation of the 
thermal surface diffusion.

\section{Results}
\label{sec:results}

We present the results of our simulations in two subsections. Before
showing the results for the growth simulations with the KMC--MD method
in section~\ref{sec:resultskmcmd},
we present and discuss MD simulations of single atom--surface collisions
in section~\ref{sec:resultsmd} below.
We start out discussing the choice of system size in the MD simulations,
and how for small systems this affects the outcomes of the collisions.
We then move on to classify and quantify the collision--induced
processes influencing atomic mobility found
at relatively low energies ($<35 eV$) for Ag$\rightarrow$Ag(111) and 
Pt$\rightarrow$Pt(111). 
Unless otherwise specified, the collisions occur at normal incidence and
with random impact parameters.

\subsection{MD of single atom--surface collisions}
\label{sec:resultsmd}

\subsubsection{Choice of MD system size: avoiding reflected energy from
the boundaries}
\label{sec:resultsmdsize}

\begin{figure}[htbp]
  \begin{center}
    \leavevmode
    \epsfxsize=8cm
    \vspace*{3mm}
    \epsffile{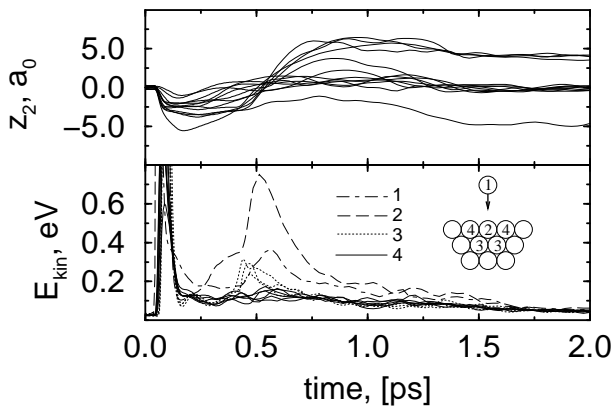}
    \vspace*{3mm}
    \caption{{\bf False energy reflections from the MD simulation boundaries:}
          25 eV Ag$\to$Ag(111) impacts. The lower panel
      shows the  average over many 
          impacts of the kinetic energy of individual atoms at various
          positions as a function of time after the impact. The inset
          schematically shows the different types of 
          positions. 1 is the incoming atom and 2 is the impact
          atom. In the fcc lattice, there are really 3 atoms in the
          position labeled 3, and 6 atoms in the position labeled 4.
          The system has $\bar{N}=(3,4,3,4)$ and
          $\bar{\xi}=(0.005,0.010,0.015)$. 
          (Same as $S_1$ in
          Fig.\protect\ref{fig:avekin}). The reflected energy from the
          boundaries is evident. After an initial peak around 0.1 ps
          the atoms slow down. Then around 0.4 ps, atoms at position
          3 (dotted curves) speed up, and immediately after that the impact
          atom accelerates wildly. The upper panel shows the vertical
          position $\rm z_2$ of the impact atom for thirteen
          individual impacts. Starting out in the surface layer at
          $\rm z_2 = 0$, the impact atom gets kicked towards the
          surface, where it bounces off the second layer and moves
          back out. Around 0.5 ps its fate is determined. It either
          stays in the surface layer, or pops out by one atomic plane
          ($\rm z_2 \to 4$);
          in a single case it squeezes into the second layer ($\rm z_2 \to -4$).          The reflected 
          energy assists the impact atom in popping out.
          }
    \label{fig:eref_pop}
  \end{center}
\end{figure}

In atom--surface collisions, the deposited energy is dissipated
away from the local impact region by shock--waves and
phonons. Clearly, how much energy stays around for how long a time is
crucial for how much atomic mobility the collision causes. When doing
a cluster simulation of the collision, it is possible that the finite
system size will affect the energy density in the impact region after
the collision --- outgoing energy waves will be reflected at the
boundaries and come back to the local impact region. This is
illustrated in figure~\ref{fig:eref_pop}, which shows results for 
25 eV Ag$\to$Ag(111) impacts for a system setup given by 
$\bar{N}=(3,4,3,4)$  and Langevin coefficients 
$\bar{\xi}=(0.005,0.010, 0.015)$. 
The lower panel in the figure shows the kinetic energy averaged
over many impacts
for individual atoms near the impact site as a function of time. 
For the first 0.4 ps, this
graph is indistinguishable from the corresponding graph for a much
bigger system. At 0.4 ps, we see that the second--layer atoms (\#3) suddenly
accelerate, and immediately after that the impact atom in the first
layer (\#2) gets a boost of kinetic energy. This sudden
acceleration after 0.4 ps is absent in a calculation for a
sufficiently big system, and is due to energy reflected by the static
atoms at the boundaries of the cluster. The upper panel in the same
figure shows the height of the impact atom $z_2$ above the
first surface--layer as a function of time,
for thirteen individual impacts. We see that this atom
sometimes stays in the same layer, sometimes pops out of the surface
layer and equilibrates as an adatom, and once ends
up one atomic layer down. The fact that the impact atom can pop out
is an energy--induced effect. It does not happen for thermal
impacts. However, the figure illustrates that for
system sizes this small, the impact atom gets a sizeable boost of energy
reflected from the boundaries right at the critical time: just where some
trajectories take the atom out of the surface layer and some
don't. And, as we shall see shortly, the probability of surface atoms
popping out in the energetic impacts depends on the system size for
small clusters. 

\begin{figure}[htbp]
  \begin{center}
    \leavevmode
    \epsfxsize=8cm
    \vspace*{3mm}
    \epsffile{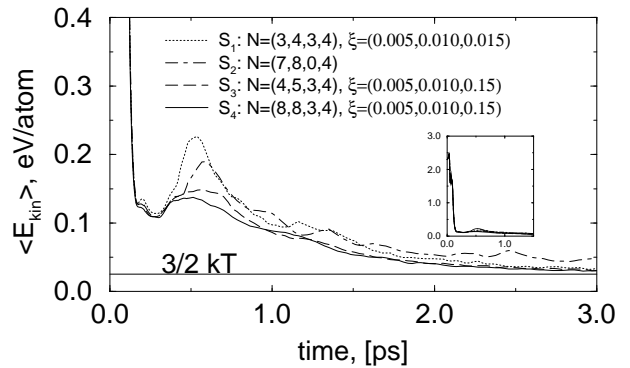}
    \vspace*{3mm}
    \caption{{\bf System size and Langevin reduction of energy reflection
        artifact.}
        Average kinetic energy of the 11 local atoms at
    positions 1 to 4 in Figure~\protect\ref{fig:eref_pop} for one
    hundred 25 eV Ag $\to$ Ag(111) impacts in 4 different system setups. The
    initial surface temperature is 17 meV (200 K). The inset (same units as
    the main plot) is the same plot on a different scale, showing the
    initial drop in kinetic energy as the 25 eV is distributed over
    many degrees of freedom. The main plot focuses on a bump in the
    local kinetic energy observed after 0.5 ps, and shows the system
    size dependence of this bump. For small systems, the bump is too
    high due to energy reflected from the boundaries.
    The setup $\rm S_2$ has no Langevin atoms.
    The setup $\rm S_3$ has the same size as
    $\rm S_2$,  but $\rm S_3$ has 3 shells of
    Langevin atoms, and the damping clearly reduces the amount of reflected
    energy.
    }
    \label{fig:avekin}
  \end{center}
\end{figure}

Figure~\ref{fig:avekin} shows the kinetic energy of the 11 atoms
neighboring the impact site (average of the 11 curves in the lower
panel of figure~\ref{fig:eref_pop}), again averaged over many impacts,
and now for four different system sizes and Langevin coefficients. 
This plot shows a bump in the
local kinetic energy around 0.5 ps after the impact, and that the size
of this bump depends on the system size and on the Langevin
coefficients of the boundary atoms. The small system setup $S_1$ is
the same as in figure ~\ref{fig:eref_pop} and has a total of 757
dynamic atoms (MD--atoms + Langevin atoms), $S_2$ and $S_3$ both have
1093 dynamic atoms (for $S_3$ some of these are Langevin atoms), and
$S_4$ has 2741 dynamic atoms.  As noted above, curves for
smaller systems can deviate from those for bigger system sizes
because of reflections
from the boundary. That the curve for $S_1$ deviates
significantly from $S_4$ after 0.36 ps, whereas $S_2$ and
$S_3$ deviate from $S_4$ only after 0.45 ps, reflects the
fact that that the bigger
the system the later the traveling reflected energy wave will affect
the local impact region. We believe that the bump in the curve for $S_4$
will not disappear as the system size is increased further; extrapolating
from the times at which the curves for $S_1$, $S_2$, and $S_3$ deviate from
$S_4$, we estimate that the
reflected energy for $S_4$ does not return until after roughly 0.5 ps,
at which time this bump has already reached its peak.
% The bump in the curve for $S_4$ is real: the
% curve is converged at least until roughly 0.5 ps, at which
% time the bump peaks.
The size of the bump is reduced from $S_1$ to
$S_2$ by increasing the system size. It is then reduced further from
$S_2$ to $S_3$ merely by tuning the Langevin coefficients on the
boundary atoms. We have done an extensive numerical exploration to
find the optimal choice of Langevin coefficients, and found 
$\xi_1 = 0.005$, $\xi_2 = 0.010$, and $\xi_3 =  0.15$ to be a good
choice. A Langevin coefficient too high and too close to the impact
region will cause energy reflection at early times directly from
these Langevin atoms. The value $\xi_3 =  0.15$ seems to be the
optimum for the outer most layer of Langevin atoms. We have also tried
random Langevin coefficients within each layer, to avoid focusing of
the reflected energy, but without finding improvements. 
It is evident that the curve for $S_3$ is not completely converged:
the local kinetic energy is affected by the boundaries for
the system $S_3$. However, we shall argue below that it is
sufficiently converged that it does not significantly affect the atomic
rearrangements in which we are interested. 

We note that the time at which the reflected energy
comes back tells us that the energy wave travels at a supersonic speed.
This is due to anharmonicities in the potential energy at the relatively 
large atomic displacements: the atomic collisions at the earliest times
are presumably
exploring the hard core of the potential and hence can propagate faster than
harmonic sound waves. Potential energy anharmonicities
({\it e.g.,} backscattering from hard collisions) can also 
cause reflection of outgoing energy even from the MD--atoms following the true 
dynamics.
We believe these anharmonicities cause the bump in
local kinetic energy found around 0.5 ps in figure~\ref{fig:avekin} for
the largest system (curve $S_4$).

\subsubsection{Adatom/vacancy formation}

\begin{figure}[htbp]
  \begin{center}
    \leavevmode
    \epsfxsize=8cm
    \vspace*{3mm}
    \epsffile{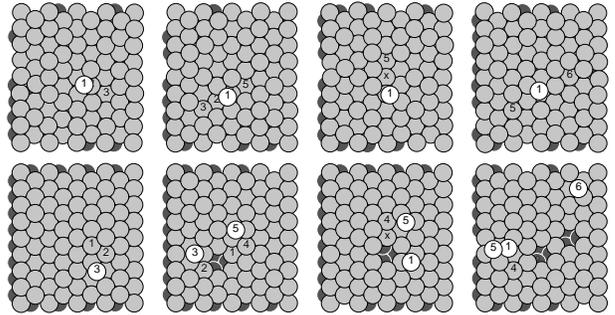}
    \vspace*{3mm}
    \caption{Possible outcomes of 35 eV impact of Ag $\to$
    Ag(111) initially at 200 K. The top row shows initial configurations
    with atom 1 (the incoming atom) above the 
    surface, and the bottom row the corresponding final
    configuration after thermalization. Atoms shifting lattice site 
    are numbered.
    The three right columns show examples of adatom/vacancy
    formation, while the left column is an example of an exchange event
    with no net effect on the substrate. The processes shown in the
    first and second column each account for about 30\%\ of the
    impacts at 35 eV. The process in the third column is an
    exchange event in which four neighbors of the atom marked "x"
    forming a chain in the first and second layers each shift positions
    by one atomic distance. (Unlike earlier figures, here ``x'' does not
    mark the impact atom, which in fact occupies the site where the vacancy
    is found after the collision.)
    At 25 eV, the process in the first column is the single
    most common event; and the process shown in the second column is
    the most frequent way of forming an adatom/vacancy pair. Projected
    trajectories for this process are shown in
    Figure~\protect\ref{fig:e35imptraj}. 
    }
    \label{fig:exage35imp}
  \end{center}
\end{figure}

Figure~\ref{fig:exage35imp} shows possible atomic rearrangements
that result from 35 eV Ag $\to$  Ag(111) impacts on an initially flat
surface. In one case (first column from left), there is an exchange
process, in which the incoming atom gets incorporated into the surface
layer, and a surface atom pops out. The impact has no net effect on the
substrate. The only net effect of the incoming energy is that the resulting
adatom ends up a couple of atomic spacings away from the impact
site. It is an example of energy--induced short ranged horizontal
mobility, which is very common for these low energy impacts. In the
second column from the left, something more happens. The incoming atom
gets incorporated into the surface, but two surface atoms pop out. 
Compared to a thermal deposition, which would result in one adatom on
the surface, the energetic impact causes the formation of an
additional adatom/vacancy pair.  This can potentially have a
significant influence on the subsequent growth. Since these two
adatoms are created very close to each other, they have a
large likelihood of meeting and forming a dimer. We shall return to this
point in section~\ref{sec:resultskmcmd}. Figure~\ref{fig:exage35imp}
also shows an example in which two surface vacancies are formed, and
a total of three atoms end up outside the surface layer. Two of these
atoms have met to form a dimer during the collision process. 

% XXX Formatting for double column
\vfil\break

\begin{figure}[htbp]
  \begin{center}
    \leavevmode
    \epsfxsize=8cm
    \vspace*{3mm}
    \epsffile{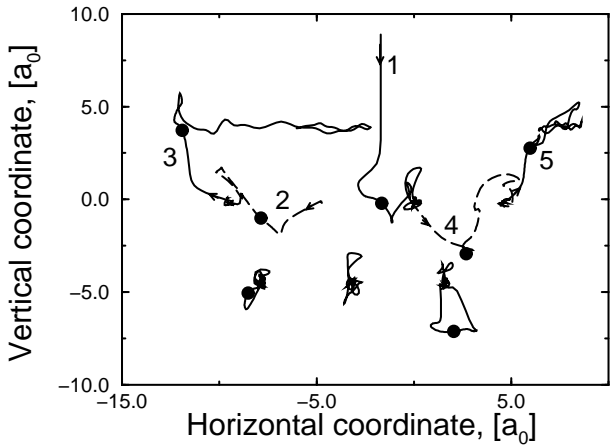}
    \vspace*{3mm}
    \caption{Projected trajectories for the adatom/vacancy forming
    process in the second column of
    Figure~\protect\ref{fig:exage35imp} (one of the 35 Ag $\to$ Ag(111)
    impacts). The
    horizontal coordinate is the one along the atomic row of atoms 3, 2, 4
    and 5 in their initial position. We see atom 1 coming in and hitting 2
    and 4 (dashed lines), knocking them down and to the sides. 2 and
    4 bounce off 
    the second layer and, still moving outwards, they push 3 and 5 out
    of the surface and take their places. The dots show the atomic
    positions after 0.25 ps. This is an example of a 35
    eV trajectory, but it is also by far the most frequent mechanism
    of forming adatom vacancy pairs at 25 eV. How far atoms 2 and 4
    move down may vary --- the main feature is that the incoming
    atom pushes two surface atoms apart, and these then each push a
    neighbor out of the surface layer.}
    \label{fig:e35imptraj}
  \end{center}
\end{figure}

Before we turn to a discussion of the probability of making these
adatom/vacancy pairs as a function of the incoming energy, let us look
into the mechanism by which they are
formed. Figure~\ref{fig:e35imptraj} shows the atomic trajectories for
the impact shown in figure~\ref{fig:exage35imp} column 2, in a vertical
cut. This process is the most frequent way of forming adatom/vacancy
pairs at 25 and 35 eV. We see atom 1 coming in, squeezing atoms 2 and
4 down and apart. After bouncing off the atoms in the second surface
layer, the atoms 2 and 4 push atoms 3 and 5 out of the surface. Atom
1 takes the place initially occupied by atom 4. The picture that
emerges is very ballistic. A collision sequence and ballistic
reflection from the deeper lying atoms cause surface atoms to pop
out. 

\begin{figure}[htbp]
  \begin{center}
    \leavevmode
    \epsfxsize=8cm
    \vspace*{3mm}
    \epsffile{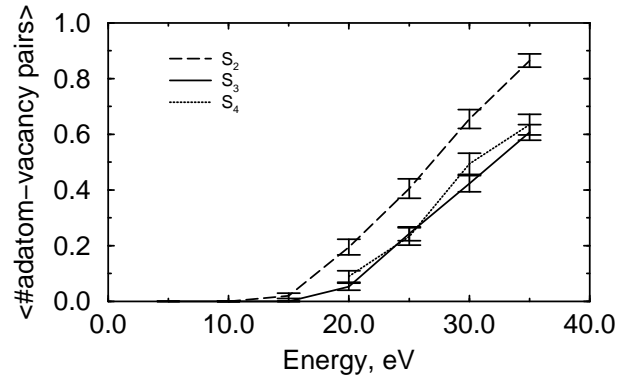}
    \vspace*{3mm}
    \caption{Average number of adatom--vacancy pairs created in
      Ag$\to$Ag(111) impacts on an initially flat surface at 200 K as
      a function of the impact energy. Each data point is based on
      200-300 hundred simulated impacts, and the error bars give the
      statistical uncertainty.  The 3 setups $\rm S_2$-$\rm S_4$ are defined in
      Figure~\protect\ref{fig:avekin}. 
}
    \label{fig:agvacform}
  \end{center}
\end{figure}

Figure~\ref{fig:agvacform} shows the average number of adatom/vacancy
pairs formed due to energetic Ag$\to$Ag(111) impacts, as a function of
energy, and
for the three system setups $S_2$, $S_3$ and $S_4$ discussed above.
The production of adatom/vacancy pairs is too high in the relatively small
system with no Langevin atoms $S_2$, and we have discussed above how
this is due to energy reflected from the system boundaries coming back
to the local impact region. However, by introducing the shell of Langevin
atoms as in system $S_3$, we find an
adatom/vacancy production which agrees within the error bars with that
of the significantly bigger system $S_4$. Based on this, unless
otherwise stated, in the following we use systems at least as large as
$\rm S_3$,
{\it i.e.,} $\bar{N}=(\ge 4,\ge 5,3,4)$, with Langevin coefficients 
$\bar{\xi}=(0.005,0.010, 0.15)$. $N_1$ and $N_2$ can be increased
at higher energies.

Figure~\ref{fig:agvacform} shows a significant production of
adatom/vacancy pairs for Ag$\to$Ag(111) impacts with incoming energies
greater than 20 eV. We shall see how this will influence the growth of
the surface in section~\ref{sec:resultskmcmd}. We have tested how this
adatom/vacancy production changes when we change the angle of incidence of the
energetic beam. For 20 eV and 25 eV Ag/Ag(111) impacts, at an angle 30
degree off normal incidence, we find average adatom--vacancy
productions per impact of $0.1 \pm 0.02$ and $0.23 \pm 0.03$,
respectively. This calculation was done in a large cell with $N_1 =
8$,$N_2 = 6$.  Thus, at this angle, the adatom/vacancy production is
essentially the same as that for normal incidence.

\begin{figure}[htbp]
  \begin{center}
    \leavevmode
    \epsfxsize=8cm
    \vspace*{3mm}
    \epsffile{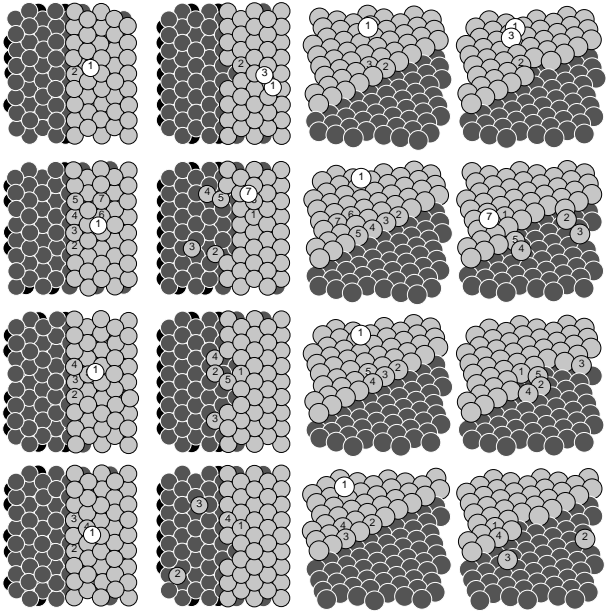}
    \vspace*{3mm}
    \caption{Possible outcomes of 25 eV impacts above a straight step on
    Ag(111) initially at 200 K. From the left the columns are top view,
    initial and final configuration, then a view from an angle, initial
    and final configuration. The atoms which shift lattice sites are
    numbered, with 1 as the incoming atom. (The distance to the step is
    between -1 and -2 on Fig.~\protect\ref{fig:agstepimp}.)
    From the top row down we see
    examples of 1) the equivalent of adatom/vacancy formation
    resulting in a dimer above the step; 2) no interlayer mobility,
    but a change of the step structure; 3) and 4) a net downward
    mobility by insertion of the incoming atom, as well as a change of
    the step structure. 2) and 4) also show the breaking off of adatoms from
    the step.}
    \label{fig:exagstepimpm1} 
  \end{center}
\end{figure}

\subsubsection{Atom insertion}

We now proceed to examine the kind of energy--induced atomic
rearrangements found for impacts near straight steps on the surface. 
We find it important to look at possible outcomes
of these single impacts in order to identify and categorize the energy--induced
atomic processes influencing the growth. On the other hand, it will
soon become evident that it is not feasible to quantitatively
describe all possible outcomes of impacts on all possible atomic
configurations of the surface. To see how energetic impacts may
influence the total interlayer mobility during the growth we look at
impacts in the vicinity of one of the two possible straight steps on
the fcc(111) surface: the so--called B-step exposing a (111) micro
facet at the step edge.

Figure~\ref{fig:exagstepimpm1} shows a selection of possible outcomes
of 25 eV Ag$\to$Ag(111) impacts just above a straight B-step. The
first row shows the production of an adatom/vacancy pair, equivalent to
what can take place on the flat surface, as discussed above. In this
particular case the two resulting adatoms on the upper terrace have met
to form a dimer in the first few picoseconds after the impact. The
second row shows an example of no net interlayer mobility due to the
impact. The incoming atom gets incorporated into the upper terrace, but another
atom (labeled 7) pops up. However, the impact does change the step
structure, which is no longer straight. The collision has produced
kinks on the step, and the atom labeled 3 ends up detached from the
step. This can influence the growth, because it might change the
subsequent thermal diffusion. For example, the thermal interlayer
mobility might be different at steps with kinks~\cite{kmcpt2d3d}. 

Figure~\ref{fig:exagstepimpm1} rows 3 and 4 show examples of an 
energy--induced net downward interlayer mobility. Where for thermal impacts
the incoming atom would stay on top of the upper terrace, it now gets
inserted into the upper terrace due to the energy. This kind of process has
previously been seen in molecular dynamics simulations of
Pt$\to$Pt(111) impacts at similar energies~\cite{villarba}. Imagine the
step is surrounding an island on the surface. The effect of the energy
in this case is to change vertical growth of the island to horizontal
growth at its periphery, and hence this insertion process will favor
smooth growth of the surface. The examples show that the insertion
process also causes a change of the step structure, and possibly
detachment of atoms from the step to the lower terrace.

% XXX Formatting for double column
\vfil\break

\begin{figure}[htbp]
  \begin{center}
    \leavevmode
    \epsfxsize=8cm
    \vspace*{3mm}
    \epsffile{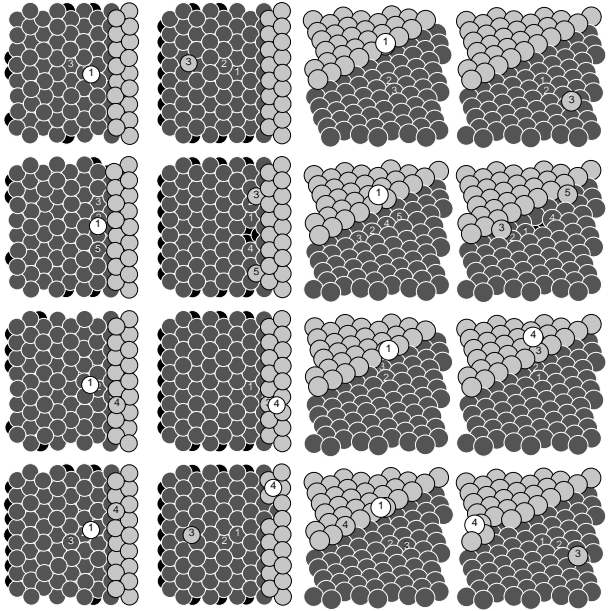}
    \vspace*{3mm}
    \caption{Possible outcomes of 25 eV impacts below a straight step on
    Ag(111) initially at 200 K. From the left the columns are top view,
    initial and final configuration, then a view from an angle, initial
    and final configuration. The atoms shifting lattice site are
    numbered, with 1 as the incoming atom. (The distance to the step is between
    1 and 2 on Fig.~\protect\ref{fig:agstepimp}.)
    From the top row down we see  examples of: 1) an exchange event
    with no net change of the substrate; 2) an adatom/vacancy
    formation resulting in attachment to the step; and 3) and 4) net
    upward mobility where a step atom gets piled up on the island.
    }
    \label{fig:exagstepimpp2}
  \end{center}
\end{figure}

\subsubsection{Atom pile--up}

Figure~\ref{fig:exagstepimpp2} shows examples of possible outcomes
of 25 eV Ag$\to$Ag(111) impacts just below a straight B-step. The
first row shows an impact which results in an exchange process involving
two surface atoms and no net interlayer mobility. The only net
effect of the energy is a short ranged horizontal displacement of the
adatom from the impact site, as we have seen also happens for impacts 
on flat parts of the surface. The second row shows an example of an
adatom/vacancy formation on the lower terrace. This process is the
same as the one for which the atomic trajectories are shown in
figure~\ref{fig:e35imptraj}, only here the two adatoms can attach to
the ascending step.

Figure~\ref{fig:exagstepimpp2} rows 3 and 4 show examples of an energy
induced net {\it upward} interlayer mobility. The atomic
trajectories for the process in row 3 are very similar to those for
row 1, only now atom 3 pops up in the step edge, and the step atom 4
gets piled up on the upper terrace. If the
step is surrounding an island, this pile--up process causes vertical
growth of the island, and hence it favors a rougher growth.

\begin{figure}[htbp]
  \begin{center}
    \leavevmode
    \epsfxsize=8cm
    \vspace*{3mm}
    \epsffile{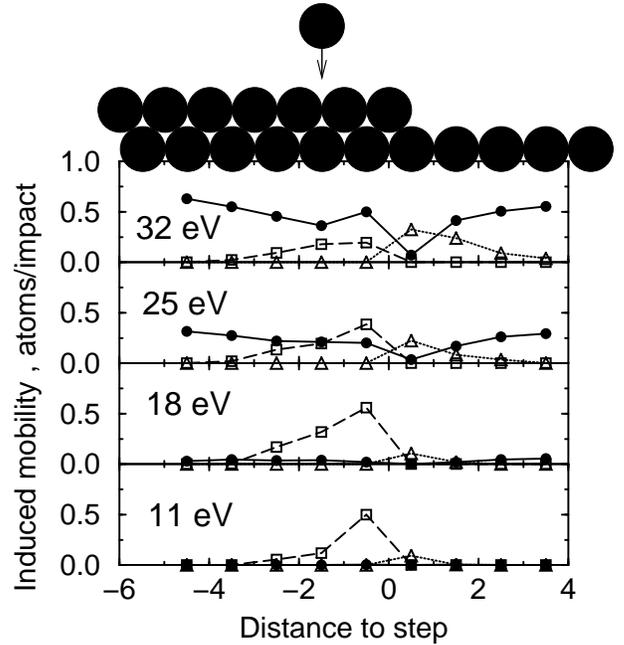}
    \vspace*{3mm}
    \caption{Impact-induced interlayer mobility as a function of the
    distance to a straight B-step for Ag$\to$Ag(111) (200K) at different
    energies. We show the average number of adatom/vacancies formed
    (filled circles, Fig.~\protect\ref{fig:exagstepimpm1} row
    1, Fig.~\protect\ref{fig:exagstepimpp2} row 2); the probability of
    insertion events (open squares,
    Fig.~\protect\ref{fig:exagstepimpm1} rows 3 and 4); and the average
    number of atoms piled up on the island (open triangles,
    Fig.~\protect\ref{fig:exagstepimpp2} rows 3 and 4).
    The ``distance to step'' is the horizontal distance in lattice constants to
    to the center of the step edge atoms (as shown at the top).
    We average over impact
    parameters within the same atomic row parallel to the step, and
    plot the result at the half integer distance (impact parameters
    between 0 and 1 are averaged over, and plotted at 0.5
    etc.). Statistical errors are less than 4\% .
}
    \label{fig:agstepimp}
  \end{center}
\end{figure}

\subsubsection{Dependence on energy and step position}

Let us now turn to a quantitative discussion of the probabilities of 
the various interlayer processes we have seen can happen for impacts 
near straight steps. Again we look at impacts near the straight B--step,
and emphasize that this is merely one example of many surface configurations 
an incoming energetic atom can encounter.
Figure~\ref{fig:agstepimp} shows the average net number of atoms per 
impact moved up or down by adatom/vacancy formation, insertion, or 
pile--up as a function of the distance to the step, and for 4 different  
incoming energies in the case of Ag$\to$Ag(111). At 11 eV the only 
significant energy--induced process is the insertion mechanism. For impacts  
between the first and the second row of atoms in the upper terrace,
there is roughly a 50\%\ chance of having an insertion event, hence 
there is a downward mobility of roughly 0.5 atom per impact. For impacts
between the second and the third row of atoms, this probability has decreased
to approximately 10\% . There is a few percent chance of insertion 
events for impacts between the third and the fourth atomic row in the upper
terrace. 
Pile--ups are defined to be when an adatom ends up on the upper 
terrace for impacts outside the crystal atomic position of the step atoms,
{\it i.e.,} for ``distance to step'' greater than zero.
At 11 eV
we see that there is a few percent chance of pile--ups for impacts within
one atomic row width from the step. We note that with this definition, 
it is possible that there would be a few pile--ups even 
thermally for impacts very close to the step edge.

At 18 eV, the picture is more or less as at 11 eV, except that there is an
increased probability for the insertion events, in particular for impacts
between the second and fourth atomic row in the upper terrace. We also see
a few percent chance of adatom/vacancy production. The increase in the
insertion events does not continue with increasing incoming energy. 
At 25 eV it is slightly reduced compared 18 eV, and at 32 eV is further
reduced as adatom/vacancy production becomes more dominant.

But the number of pile--up events has increased with incoming 
energy, and at 32 eV, there are more pile--up events for impacts below
the straight step than insertions for impacts above the step. Both
pile--ups and insertions happen within the width of three to four atomic
rows from the step. In agreement with figure~\ref{fig:agvacform} the
average number of adatom/vacancy pairs per impact has increased at 25
and 32 eV to approximately 0.3 and 0.5 for impacts not too close to the
step. However, there is a clear dip in the probability for these events
close to the step, in particular for impacts right below the step.

Based on figure~\ref{fig:agstepimp}, one might conclude that growing
a Ag(111) surface using a low energy atom or ion--beam might have an 
advantageous effect on the surface roughness, since at 18 eV we see 
a significant energy--induced downward mobility near step edges. 
However, one would also expect an optimal energy for growing a smooth surface,
since at a little higher energy the total pile--up exceeds the total
insertion at the straight step. This is indeed what we will see in 
section~\ref{sec:resultskmcmd} in a certain growth parameter range.
However, it must 
be kept in mind that the importance of insertions and pile--ups will
depend on the step density on the surface, and that when the steps on the 
surface are rough the probabilities for insertions and pile--ups
might be different from those for the straight step.

\subsubsection{Pt$\rightarrow$Pt(111) impacts}

\begin{figure}[htbp]                                            
  \begin{center}
    \leavevmode
    \epsfxsize=8cm
    \vspace*{3mm}
    \epsffile{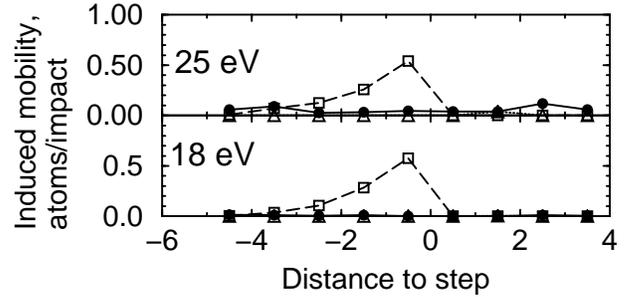}
    \vspace*{3mm}
    \caption{Impact--induced interlayer mobility as a function of the
    distance to a straight B-step for Pt$\to$Pt(111) (200K) at different
    energies.  We show the average number of adatom/vacancies formed
    (filled circles); the probability of
    insertion events (open squares); and the average
    number of atoms piled up on the island (open triangles) --- see
    text to Figure~\protect\ref{fig:agstepimp}.
}
    \label{fig:ptstepimp}
  \end{center}
\end{figure}

Figure~\ref{fig:ptstepimp} shows the results of the same kind of
calculation presented in figure~\ref{fig:agstepimp}, but now in the
case of 18 eV and 25 eV Pt$\to$Pt(111) impacts. The energy--induced
processes in the vicinity of the step are predominantly insertions. 
Both at 18 eV and 25 eV, there is a significant probability of
insertion events up to three to four atomic rows from the step. There
are no pile--up events at these energies. At 25 eV there is a few percent
chance of
producing adatom/vacancy pairs during the impact and, by comparison to
Ag$\to$Ag(111), we conclude that this process must have an onset at a
higher energy. 

\begin{figure}[htbp]
  \begin{center}
    \leavevmode
    \epsfxsize=8cm
    \vspace*{3mm}
    \epsffile{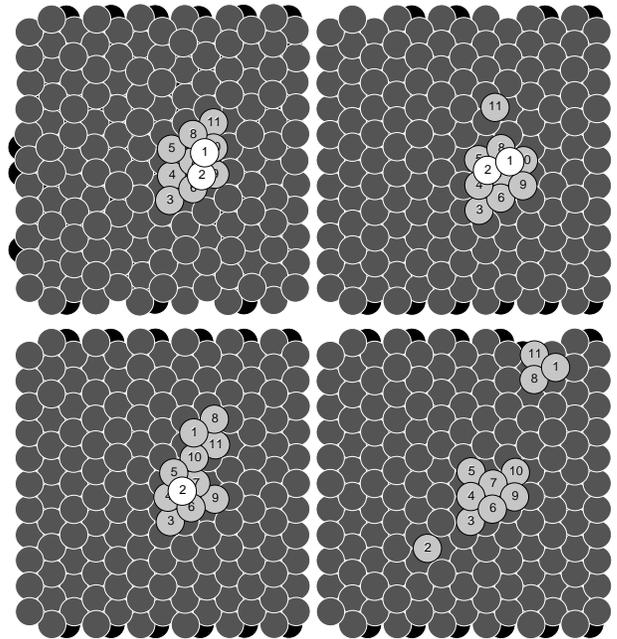}
    \vspace*{3mm}
    \caption{Results of 18 eV Ag$\to$Ag(111) impacts on a small
    island initially with an adatom atop. The initial configuration is
    shown in the upper left --- the incoming atom (labeled 1) is a few
    atomic distances above the island. Three possible outcomes are: 1) two
    atoms atop the island (upper right); 2) one atom atop the island
    (lower left); and 3) no atoms atop the island (lower right).}
    \label{fig:atomonisl}
  \end{center}
\end{figure}

\subsubsection{Insertion of additional atoms and island break--up}
\label{sec:atomatop}

Another mechanism favoring smooth growth was revealed to us
watching movies of the full growth from the KMC--MD simulations. 
In impacts on a small island with an adatom atop, it is possible
that the incoming atom and the adatom initially atop the island both
end up in the layer of growing island. Thus, not only is the incoming
atom inserted into the island, an existing adatom on the island is
inserted as well. In these
processes the small island is likely to undergo structural changes,
perhaps even break up. We tried thirty 18 eV Ag$\to$Ag(111) impacts
on a 3 by 3 atom island with an adatom initially on top, with random
impact parameters in a surface unit cell neighboring the top
adatom (see fig.~\ref{fig:atomonisl}). Of these 30 impacts, both atoms ended
up in the growing  layer in 19 cases (2 atoms down) , one atom stayed
on top of the island in 7 cases (1 atom down), and in 3 cases both
atoms stayed on the island (no induced interlayer mobility). In 
one case 3 atoms ended up on top of the island (1 atom
up). Regarding the impact induced break--up of the island, only in
8 cases did the 11 atoms end up as one connected island, while in 22 cases
one or more atoms ended up separated from the original island.

\subsubsection{Summary of MD simulations of single impacts}
\label{sec:resultsmdsum}

In summary we can categorize the collision--induced events as follows:

\begin{itemize}
\item {\bf Adatom/vacancy formations}. The impact causes the
  digging up of atoms from the surface layer.
\item {\bf Insertions}.  Near descending steps, the
  incoming atom can be incorporated into the upper terrace. Insertion can also
  happen for impacts on small islands.  Adatoms initially residing
  above descending steps can also be inserted due to an energetic impact.
\item {\bf Pile--ups}.  Near ascending steps,
  impacts below the step can result in net growth on the upper terrace.
\item {\bf Step edge restructurings}.  Impacts near steps can change the
  step edge structure.
\item {\bf Island break--ups}.  Impacts on small islands can cause the
  island to break up into several pieces. Also, impacts near straight
  steps can result in adatoms detaching from the step and ending up
  isolated. 
\item {\bf Horizontal mobility}.  Incoming atoms may be displaced
  horizontally before thermalizing. Also, all the events above involve
  some induced horizontal mobility.
\end{itemize}

Some of these events must be characterized as energy--induced defect
formations, leaving the surface in a higher energy state when compared
to a thermal deposition. Adatom/vacancy formations, pile--ups and 
island break--ups are such examples. Insertions, on the other hand, 
can take the surface to a lower energy
state, by incorporating the incoming atom or other adatoms initially 
residing above descending steps. In this respect insertions are examples of 
energy--induced annealing of defects. However, as we have seen,
insertions may be accompanied by an energy increasing restructuring
of the step edge, by creating step edge defects like kinks. Also, 
step edge restructurings are defect formations for impacts on
initially straight steps, but for impacts near rough steps, they could
be annealing events. 

One of the questions one might hope to answer by doing these molecular
dynamics simulations of single impacts is what the nature of the
energy--induced mobility is. Does the energy, for example, cause a local
heating, and can we then understand the increased mobility to be a result of
a locally higher temperature? Or is the impact induced mobility
more ballistic in
nature? Based on the results presented here, we conclude that for
these low--energy ($\le 32$eV) metal--on--metal impacts, the
local heating and increased temperature is the wrong picture. 
Plots like the one shown in figure~\ref{fig:e35imptraj} indicate that the
more relevant picture is one
of a clearly ballistic collision sequence that, depending
on impact parameter, may result in defect formation or annihilation or
both, as discussed above. 

\subsection{KMC--MD simulations of growth}
\label{sec:resultskmcmd}

After having discussed what can happen in the energetic atom--surface
collisions by presenting our molecular dynamics simulations of such
impacts, let us now turn to the results of the combined KMC--MD
simulation of the growth by energetic deposition. We begin by
looking at the effect of the energy on the submonolayer growth
of Ag/Ag(111). Subsequently we discuss the growth of several
layers, and then we discuss how the picture changes for Pt/Pt(111). 
All the presented KMC--MD simulations are done at a deposition rate
of 1 ML/s and with normal incidence.

\onecolumn
\begin{figure}[htbp]
  \begin{center}
    \leavevmode
    \epsfxsize=14cm
    \vspace*{3mm}
% XXX Smaller
%    \epsffile{fig/fig12/cov30.ps}
    \epsffile{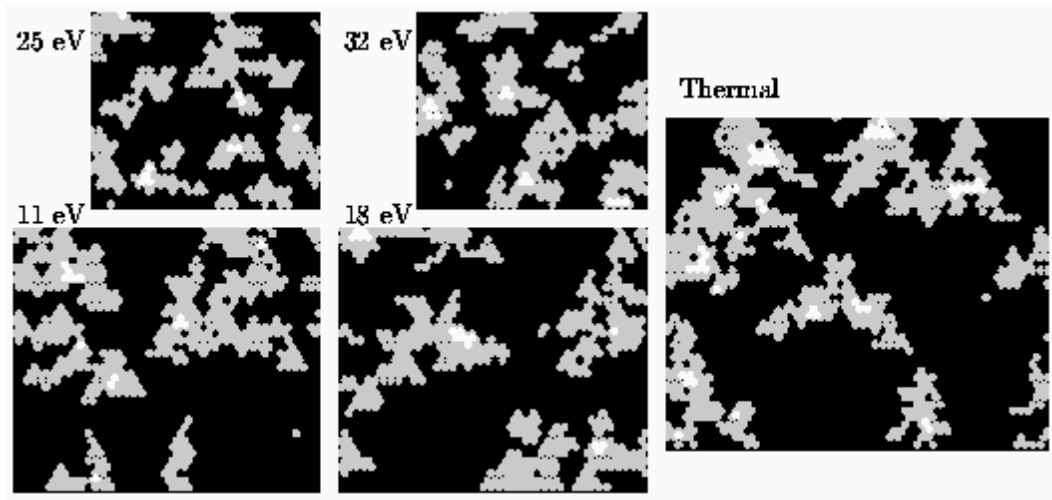}
    \vspace*{3mm}
    \caption{The surface morphology at 0.3 ML coverage for Ag/Ag(111) as 
      obtained for thermal growth and for growth by energetic deposition
        at four different incoming energies. The surface areas with
        periodic boundary
      conditions, 30$\times$30, 40$\times$40, and 50$\times$50 atomic surface 
      unit cells (decreasing with increasing energy), are chosen to give a
        minimum 
      of 4--5 islands in the simulation. First and second layer growth is
      seen in all cases. The surface temperature is 60 K.}
    \label{fig:agcov30}
  \end{center}
\end{figure}
\twocolumn

\subsubsection{Ag/Ag(111) submonolayer structure}

We have simulated Ag/Ag(111) growth at four different incoming
energies, 11, 18, 25 and 32 eV, and for thermal deposition, in which
case no MD simulations are done. For all the simulations of energetic
deposition, we simulate the atom--surface collisions using
molecular dynamics for cluster setups as explained in 
section~\ref{sec:methodmd}.
All cluster setups have Langevin coefficients $\xi = (0.005,0.010,0.15)$.
At 11 eV we use $\bar{N}=(3,4,3,4)$, 
at 18 eV and 25 eV we use $\bar{N}=(4,5,3,4)$, and
at 32 eV we use $\bar{N}=(5,6,3,4)$. We first show results where
the surface temperature is set to 60 K. This temperature determines
the rates of various diffusion processes in the KMC part of the
simulation, and is also the
Langevin temperature and the temperature to which the MD--atoms
are initially equilibrated.
We later discuss the effect of increasing the surface temperature.

Figure~\ref{fig:agcov30} shows the surface morphology at 0.3 ML
coverage for  Ag/Ag(111) at the different energies, and for the
thermal deposition.  Focusing on the thermal run, we observe island
formation due to surface diffusion. These islands are dendritic with
very irregular step edges due to limited diffusion along the step
edges at this low temperature. We also observe that the branches of
the dendritic islands have preferred growth directions perpendicular
to the so-called A-steps. This is because of the asymmetry in the {\it corner
 diffusion} in the EMT barriers, see table~\ref{tab:modbar}.
When a diffusing adatom attaches to the island at the abundant
corner--sites, they subsequently preferentially move to the A-step,
causing growth in that direction~\cite{kmcagpt}. 

\begin{figure}[htbp]
  \begin{center}
    \leavevmode
    \epsfxsize=8cm
    \vspace*{3mm}
    \epsffile{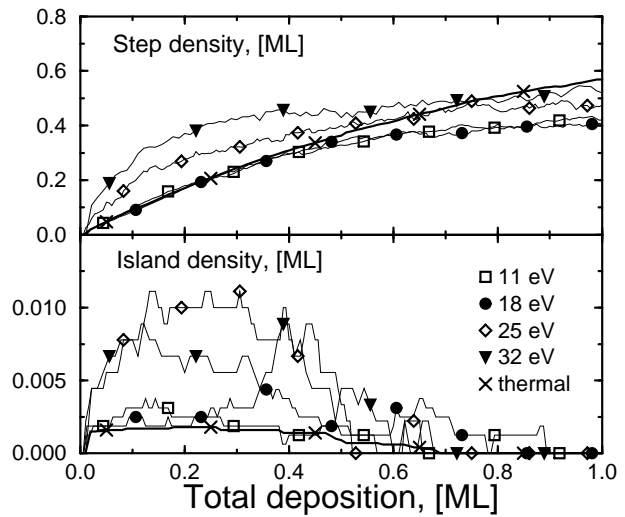}
    \vspace*{3mm}
    \caption{Island densities (lower panel) and step densities (upper
    panel), for Ag/Ag(111) at various impact energies and thermal
    deposition for submonolayer coverages. The surface temperature is
    60 K. At 25 and 32 eV the 
    island density is significantly increased compared to thermal
    deposition due to adatom--vacancy formation in impacts. This
    results in an increased step density for these energies, enhancing
    the total energy--induced insertion at steps. At 11 and 18 eV the
    island density is increased slightly
    above the thermal island density due to the breaking up of existing islands
    in impacts. Islands formed this way coalesce quickly, and the
    mechanism does not increase the step density. 
    After 0.5 layers down, the step densities for the energetic
    depositions level off due to the smooth growth, whereas the
    thermal step density continuously increases as the surface
    roughens. 
}
    \label{fig:agislstep}
  \end{center}
\end{figure}

When comparing the thermal run with the various energetic depositions
in figure~\ref{fig:agcov30}, there is one striking difference. For the
energetic depositions, in particular at 25 and 32 eV, we observe a
higher density of smaller islands. This is even more clearly shown in
figure~\ref{fig:agislstep}, where we show the island densities as a
function of coverage in the lower panel. The figure shows that for
not too high coverages the island densities during energetic
depositions are higher than that for the thermal deposition --- for 25 and
32 eV they are significantly higher. The mechanism giving rise to this increased
island density at the higher energies is the formation of
adatom/vacancy pairs in the energetic atom--surface collisions, as
discussed in section~\ref{sec:resultsmd}. Dimers are either created
directly, or they are formed with an increased probability by the two
adatoms that result from a collision--induced adatom/vacancy creation.
At 32 eV even trimers may be formed directly.
The formation of a dimer does not necessarily nucleate a new island,
since these dimers are mobile in the KMC--diffusion model, and hence
can diffuse on the surface and attach to an existing island. However,
the dimers move more slowly than single atoms --- they have a shorter
diffusion length on the time--scale of the deposition rate, resulting
in a higher island density.

\begin{figure}[htbp]
  \begin{center}
    \leavevmode
    \epsfxsize=8cm
    \vspace*{3mm}
    \epsffile{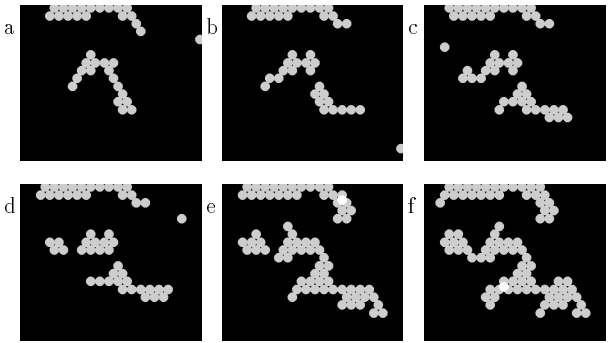}
    \vspace*{3mm}
    \caption{Evolution of an island at low coverages for 11 eV
    Ag/Ag(111) deposition at 60 K. a-f are at coverages of 0.056,
    0.063, 0.081, 0.088, 0.131, and 0.150 ML respectively. We see
    that during the deposition and the growth of the island it first
    breaks into two and then into three pieces, which coalesce again before a
    coverage of 0.15 ML. Such early coalescence can be expected for
    islands formed by the beam breaking off pieces of an existing
    island, and is generally not expected for thermal deposition.
    At the top of the images, another
    island is growing. The islands nucleated by diffusion on the
    terrace are well separated, and will typically coalesce around 
    0.5 ML coverage or later.}
    \label{fig:ag11evislbreak}
  \end{center}
\end{figure}

In section~\ref{sec:resultsmd} we found the energy onset of collision
induced adatom/vacancy formations to be around 20 eV. The lower panel
of figure~\ref{fig:agislstep} shows an increased island density even
at 11 eV. At this energy the island density is increased around 0.1
ML due to energy--induced breaking up of existing islands into smaller
pieces. Figure~\ref{fig:ag11evislbreak}a-d shows the evolution of an
island during the 11 eV energetic deposition. From a to b, the island
has grown from 17 atoms to a total of 22 atoms, but is also broken
into two pieces. The breaking up of the island is not due to thermally
activated mobility, but rather is the result of an impact on the very
narrow island. From b to c the island grows further, and then from c
to d one of the pieces again breaks into two. This break up is also caused 
by an energetic impact. The island now consists
of three pieces. And it is counted as three when the island density
plotted in figure~\ref{fig:agislstep} is calculated. From d to e two
of the pieces have coalesced by further growth, and in f at a total
coverage of 0.15 ML, the island has grown and coalesced into one
connected piece.

For thermal deposition the island density is a very useful concept,
because the average distance between islands actually is a good
measure for the typical island separation. The reason for this
is that the islands are nucleated by the random diffusion and aggregation 
of adatoms. 
In the vicinity of existing islands, adatoms diffuse and
attach at the island edge, and this gives rise to a ``denuded zone'' 
around the island with a low adatom concentration, and hence with  
strongly reduced probability of nucleating new islands. As a result 
islands on the surface tend to be approximately equally spaced, with 
a spacing determined by the diffusion of the adatoms.
As also discussed by Esch {\it et al.}\cite{esch}, this does not necessarily 
hold for energetic deposition, 
since islands can be created very close to existing ones.
Here it happens by the breaking off of pieces of existing islands.
It could also possibly happen by the direct nucleation of 
new islands by impacts on the surface near existing islands.

In the thermal case, because of the separation of the islands,
coalescence sets in at 0.4-0.5 ML coverage, and the island density
then drops down. The example in figure~\ref{fig:ag11evislbreak}
of energetic deposition shows
coalescence of the pieces of the island as early as 0.1 ML. The little
islands created by breaking off pieces of existing islands do not
survive for long. In the upper part of the pictures in
figure~\ref{fig:ag11evislbreak}a-f is a separate island which has
been nucleated independently by the diffusion of adatoms. The two
islands in figure~\ref{fig:ag11evislbreak}f are both nucleated 
by surface diffusion and behave
much like islands formed in thermal deposition. They coalesce around
0.5 ML coverage.

Also at 18 eV we see an increased island density setting in around 0.25 ML,
and by watching growth movies, we identify the mechanism to be
breaking off little pieces of the rather dendritic
islands. These little islands usually have short lifetimes before
they again coalesce with the bigger island.
At 11 and 18 eV the island density plotted in
the lower panel of figure~\ref{fig:agislstep} 
can be interpreted as the thermal
island density plus short--lived fluctuations above this level due to
breaking off pieces of islands.

At 25 and 32 eV adatom/vacancy pairs are formed in the atom--surface
collisions. Here the island density is determined first by
the fraction of impacts resulting in 1 adatom, 2
adatoms, dimers etc.\ and then the  diffusion of these species.
In addition to this, there will be relatively short lived fluctuations 
due to the breaking up of existing islands into pieces. 

In section~\ref{sec:resultsmd} we identified energy--induced
processes taking place near steps on the surface which could affect
the smoothness of the grown surface because they involve interlayer
atomic mobility. The importance of these processes is of course
determined by the total step density on the surface during the
growth. In the upper panel of figure~\ref{fig:agislstep} we plot the
step density, defined as the number of lattice sites occupied by atoms
with less than 6 neighbors in the same layer divided by the number of
sites in one layer (the unit is ML --- monolayers).  For thermal
deposition the step density can be estimated
from the island density, the coverage and a fractal dimension of the
island. For the energetic depositions, we see something different. 
The increase in island
density below 0.5 ML at 11 and 18 eV, is not reflected in an increase
in step density. The islands formed by breaking off pieces from
existing islands do not contribute to the step density in the same
way as islands nucleated by diffusion. 

At 25 and 32 eV, on the other hand, we see a significantly  increased
step density at not too high coverages. This is due to islands
nucleated from the relatively 
slow diffusion of dimers resulting from collision--induced
adatom/vacancy creations. But again, we do not have a simple relation
between island density and step density.  At 25 eV we observe
higher island density than at 32 eV, but we observe higher step density at
32 eV. While the step density is obviously an important quantity in
energetic deposition, we conclude that the island density here is a
much less useful concept than it has proven to be for thermal
deposition. 

In summary, we find that energetic deposition increases island densities and
step densities. Island densities are increased by breaking existing islands 
into pieces, and at the higher energies because of collision--induced 
adatom/vacancy formation. Only islands formed by the latter mechanism give
rise to an increased step density.

\begin{figure}[htbp]
  \begin{center}
    \leavevmode
    \epsfxsize=8cm
    \vspace*{3mm}
    \epsffile{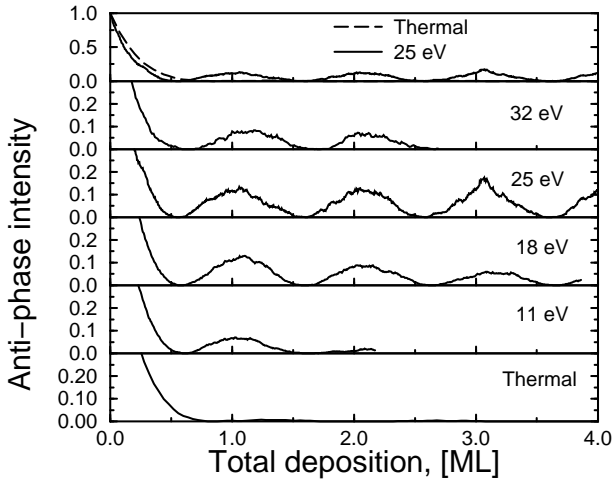}
    \vspace*{3mm}
    \caption{Simulated ``anti--phase'' intensity for Ag/Ag(111) at 60 K for
    thermal deposition, and for energetic deposition at 11, 18, 25 and
    32 eV. Energetic deposition induces oscillations indicating
    (transient) layer--by--layer growth. The strongest and
    slowest decaying oscillations are observed at 25 eV.}
    \label{fig:aglayint1}
  \end{center}
\end{figure}

\subsubsection{Ag/Ag(111) --- growth of the first few layers}

We now move on to a discussion of our results for the growth of
several atomic layers of Ag on Ag(111). Our primary focus will 
be on the smoothness of the growth, and how the smoothness is 
affected by the incoming energy. 

Figure~\ref{fig:aglayint1} shows the normalized anti--phase intensity which
would be measured in anti-phase scattering of He\cite{Rosenfeld}, 
RHEED,~\cite{RHEED}, LEED~\cite{henzler} or 
X-rays\cite{Cooper}, for our thermal and energetic deposition
simulations. The intensity $I$ is calculated from the simulations as a
function of time, using the expression
$ I = ( \sum_{i=0}^{\infty} (-1)^i (\theta_i-\theta_{(i+1)}))^2$,
where $\theta_i$ is the fractional coverage in the i'th layer. 

\begin{figure}[htbp]
  \begin{center}
    \leavevmode
    \epsfxsize=8cm
    \vspace*{3mm}
% XXX Smaller
%    \epsffile{fig/fig16/3ml.ps}
    \epsffile{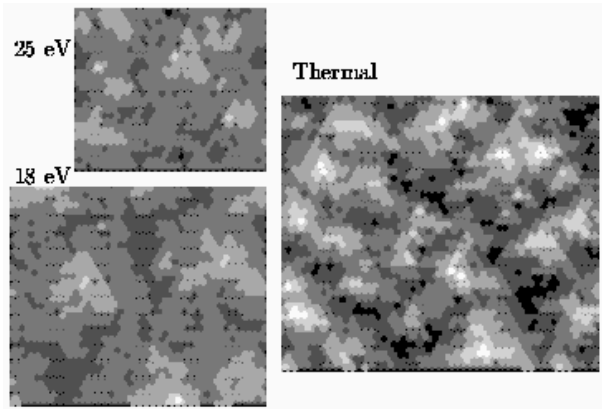}
    \vspace*{3mm}
    \caption{Surface morphologies for Ag/Ag(111) at 60 K after a
    total deposition of 3 atomic layers, for thermal (size $50\times 50$)
    growth and energetic deposition at 18 eV (size $40\times 40$) and 25 eV
    (size $30\times 30$).} 
    \label{fig:ag3ml}
  \end{center}
\end{figure}

At 60 K there is no thermally activated interlayer mobility.
Hence, for the thermal deposition, 
adatoms landing on top of existing islands will stay there. Islands are 
nucleated on top of 
islands, and multi--layer growth results. The anti--phase intensity 
$I$ drops rapidly to zero. Figure~\ref{fig:aglayint1} shows that, when the 
surface is grown by energetic deposition, oscillations in $I$ can be induced. 
The damped oscillations in $I$ correspond to a decaying layer--by--layer growth 
where the surface partly recovers its initial 
flatness after each atomic layer is deposited.

\begin{figure}[htbp]
  \begin{center}
    \leavevmode
    \epsfxsize=8cm
    \vspace*{3mm}
    \epsffile{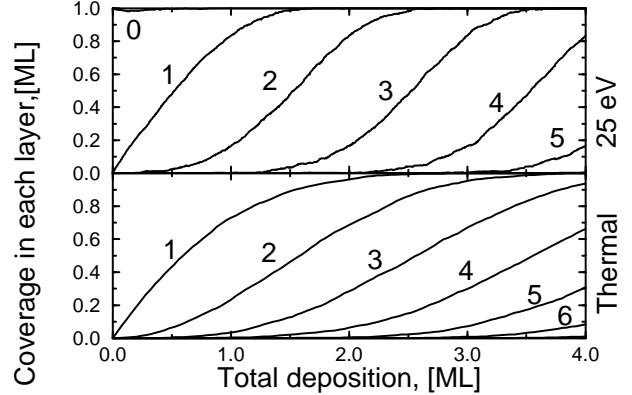}
    \vspace*{3mm}
    \caption{Plot of the coverage in each growing layer (layer \#
    labeled at each graph) above the
    initial surface as a function of the total deposition for
    Ag/Ag(111) at 60 K. Upper panel is for 25 eV, and lower is for
    thermal deposition.} 
    \label{fig:agheightdist}
  \end{center}
\end{figure}

That the surface grown by energetic deposition is more smooth than
that grown by
thermal deposition can be seen more directly from figure~\ref{fig:ag3ml}, which 
shows the surface morphology after 3 monolayers deposited, with the gray scale 
indicating the height above the initial surface, for 18 and 25 eV
and for thermal deposition. The same effect
is evident in figure~\ref{fig:agheightdist}, which shows the coverage 
in each atomic layer as a function of the total amount deposited for 25 eV and
for 
the thermal deposition. After thermal deposition of three monolayers, growth 
of the 6th layer sets in while the second layer is still not completed. At the 
same total coverage in the 25 eV deposition, the third layer is almost complete,
there is some growth in the fourth layer, and only a fraction of a percent of 
the fifth layer is occupied by atoms. 

The fact that using energetic deposition promotes the growth of flat surfaces
is consistent with our results for the energy--induced atomic 
mobilities in the atom--surface collisions presented in
section~\ref{sec:resultsmd}. 
At 60 K thermal interlayer mobility can be neglected. It is the energy--induced 
insertions  for impacts near descending steps which lead to the growth of 
more smooth surfaces --- impact parameters where the incoming
atom would stay on top of an existing island in thermal deposition but,
due to the energy, now result in lateral growth of the island.

Figure~\ref{fig:aglayint1} shows that the magnitude of the oscillations of 
$I$ gradually increase when the energy is increased from 0 eV to 25 eV. But 
by increasing the energy further to 32 eV the oscillations are again reduced. 
There is an optimal energy around 25 eV
for growing smooth surfaces. 
This is exactly what is predicted from figure~\ref{fig:agstepimp}, 
which shows the statistics of possible outcomes of impacts near straight steps. 
Figure~\ref{fig:agstepimp} shows a maximum of insertion events at 18 eV, 
pile--ups setting in at 25 eV and dominating over the insertions at 32 eV. 
Based 
on figure~\ref{fig:agstepimp}, one might have expected an optimum for growing 
smooth surfaces at 18 eV, and one might ask why we find this optimum to be 
closer to 25 eV.
Several things play a role: 1) figure~\ref{fig:agstepimp} shows the result 
for straight B-steps, but the islands at this temperature are quite small, 
and have irregular step edges. This might shift the balance of the insertion 
events to pile--ups to have the optimum at higher energies, and 2) the overall 
importance of the energy--induced events near the steps will increase with the 
step density and, as we have discussed above, the step density increases with 
increasing deposition energy after the onset of energy--induced adatom/vacancy 
formations. At 25 eV, where the insertion events still dominate over the 
pile--ups at the straight steps, the total energy--induced downward mobility
is increased due to the energy--induced increase in step density.

\begin{figure}[htbp]
  \begin{center}
    \leavevmode
    \epsfxsize=8cm
    \vspace*{3mm}
    \epsffile{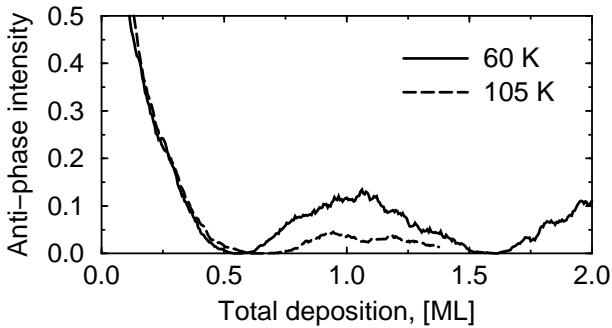}
    \vspace*{3mm}
    \caption{Simulated ``anti--phase'' intensity for Ag/Ag(111) at two 
     surface temperatures, 60 and 105 K. At the higher
    temperature the oscillations decay more rapidly, essentially due
    to a lower island density, a lower step density, and hence a lower 
    total energy--induced downward mobility.}
    \label{fig:aglayint2}
  \end{center}
\end{figure}

\subsubsection{Higher surface temperatures}

This takes us to the discussion of the effect of surface temperature 
during growth 
by energetic deposition. Figure~\ref{fig:aglayint2} shows that, for growth
at 25 eV, if we increase the surface temperature 
from 60 K to the 105 K the magnitude of the first oscillation in $I$ is reduced 
approximately by a factor of 3. At 105 K the thermal interlayer mobility is
still 
very low, but the diffusion of the adatoms on the terrace has sped up
significantly, 
reducing the island density, and hence the step density. In addition to
this the 
mobility along the island edges has increased, making islands more compact, and
reducing the step density even 
further. Thus the total energy--induced downward mobility near the step edges 
is reduced, and the surface grows rough more rapidly. It should be evident 
that the important parameter here is not the surface temperature itself, but 
rather the step density, and possibly the step edge structure. Due to the very 
low barrier for surface diffusion of adatoms on Ag(111) (which is 
underestimated in the EMT), we have to go to the low temperatures
used in our simulations to see an 
effect of the energy on the smoothness of the surface grown. 

As the temperature increases even further, the island and step
densities decrease to such an extent
that the effect of the incoming energy on the surface height
distribution is negligible. However, there can
still be a strong effect of the energy on the island density.
For example, in (slightly flawed~\cite{excuse}) KMC--MD simulations of 
25 eV Ag(111) growth at 200 K the energetic beam no longer changed the surface
height distribution, even though the island density 
changed by a factor of up to 20 compared to thermal growth.
However, we have found this to be strongly dependent
on details of the implementation of the KMC diffusion model. First of all,
the effect depends strongly on the mobility of the small clusters 
of atoms nucleated in energetic impacts. In the KMC model presented 
in this paper, these clusters are rather mobile due to periphery
diffusion of their atoms, hence reducing the energy effect on island
densities. Another subtle but important factor is the efficiency with
which diffusing dimers and trimers can fill in the isolated vacancies
created in energetic impacts. In EMT this 
readily happens. However in KMC models, where we do not allow this 
recombination, the concentration of vacancies may build up to an
extent where it hinders the diffusion of the small clusters of atoms,
and hence results in a higher island density (a point also made by
Esch {\it et al.}~\cite{esch}).

\subsubsection{Pt/Pt(111) submonolayer structure}

We now present the results of KMC simulations of thermal and 
KMC-MD simulations of 18 eV and 25 eV energetic 
deposition of Pt/Pt(111) and compare these results to the case of Ag/Ag(111).
All the Pt KMC-MD simulation are done with a surface temperature of 80 K 
and a deposition rate of 1 ML/s.
The MD part of the simulations is done using setups with 
Langevin coefficients $\xi = (0.005,0.010,0.15)$ and 
$\bar{N}=(N_1,N_2,N_3,N_4)=(4,5,3,4)$.

\begin{figure}[htbp]
  \begin{center}
    \leavevmode
    \epsfxsize=8cm
    \vspace*{3mm}
% XXX Smaller
%    \epsffile{fig/fig19/cov30.ps}
    \epsffile{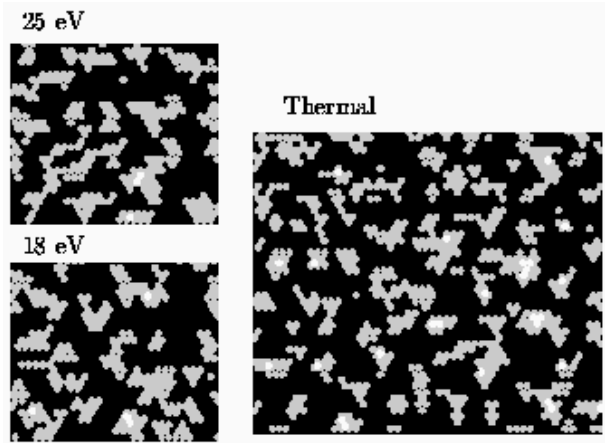}
    \vspace*{3mm}
    \caption{The surface morphology at 0.3 ML coverage for Pt/Pt(111) as 
      obtained for thermal growth and for growth by energetic depositions
        at 18 eV and 25 eV. The surface areas with periodic boundary
      conditions are 30$\times$30 and 50$\times$50 atomic surface 
      unit cells for energetic and thermal deposition, respectively.
      First and second layer 
      growth is seen in all 
      cases. The surface temperature is 80 K.}
    \label{fig:ptcov30}
  \end{center}
\end{figure}

\begin{figure}[htbp]
  \begin{center}
    \leavevmode
    \epsfxsize=8cm
    \vspace*{3mm}
    \epsffile{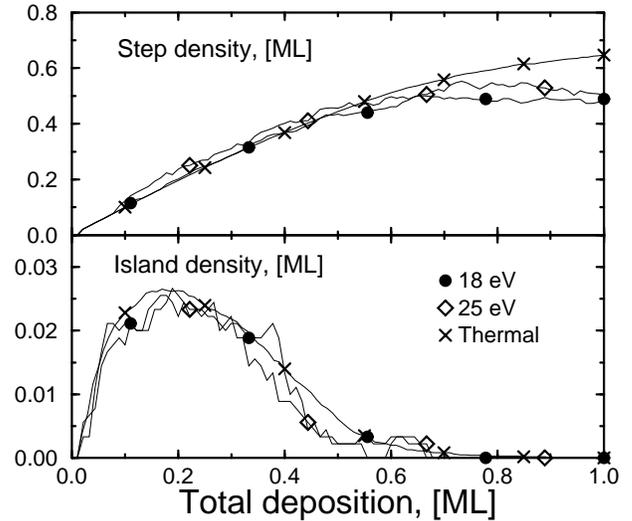}
    \vspace*{3mm}
    \caption{Island densities (lower panel) and step densities (upper
    panel), for Pt/Pt(111) at 18 eV, 25 eV and thermal
    deposition for submonolayer coverages. The surface temperature is
    80 K. In the case of Pt, 18 and 25 eV are both too low energy to 
    change the island
    and the step densities at low coverages.}
    \label{fig:ptislstep}
  \end{center}
\end{figure}

Figure~\ref{fig:ptcov30} shows the surface morphology after the deposition of
0.3 ML of Pt on Pt(111) for thermal deposition and for the two energies.
Compared to the Ag/Ag(111) simulations, there is a much higher thermal
island density. This is of course due to the higher energy barriers for surface
diffusion on Pt. Because of the higher density of islands, they are smaller and
tend to be less branched. Thus in the case of Pt, the submomolayer structure
obtained by energetic depositions at 18 and 25 eV
is very similar to the thermal submonolayer structure.
This is also apparent from figure~\ref{fig:ptislstep}, which shows the
island and step densities, as figure~\ref{fig:agislstep} did for Ag. At low
coverages the island and step densities for the energetic depositions follow 
the corresponding curves for the thermal deposition quite closely. If there
is a difference at all, the energy slightly reduces the island density.
This could be explained by short ranged energy--induced horizontal mobility.
Around 0.7 ML
the energetic step densities level off, indicating smooth growth, whereas the 
thermal one continues to increase as the surface roughens. We recall from 
figure~\ref{fig:ptstepimp} that 25 eV is below the onset for energy--induced
adatom/vacancy formations for Pt$\rightarrow$Pt(111) impacts, consistent 
with the unchanged island density. Apparently, the breaking up of existing
islands
is also not pronounced here --- probably as a consequence of the less branched 
island structure, and the stronger Pt--Pt bond energies.

\subsubsection{Pt/Pt(111) --- growth of the first few layers}

\begin{figure}[htbp]
  \begin{center}
    \leavevmode
    \epsfxsize=8cm
    \vspace*{3mm}
    \epsffile{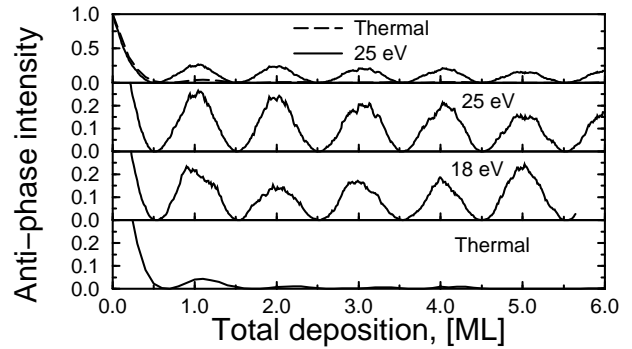}
    \vspace*{3mm}
    \caption{Simulated ``anti--phase'' intensity for Pt/Pt(111) at 80 K
    for 
    thermal deposition, and for energetic deposition at 18 and  25 eV.
    Energetic deposition induces oscillations, indicating
    (transient) layer--by--layer growth. Fairly strong and
    slowly decaying oscillations are observed at 18 and 25 eV.}
    \label{fig:ptlayint1}
  \end{center}
\end{figure}

\begin{figure}[htbp]
  \begin{center}
    \leavevmode
    \vspace*{3mm}
    \epsfxsize=8cm
    \vspace*{3mm}
    \epsffile{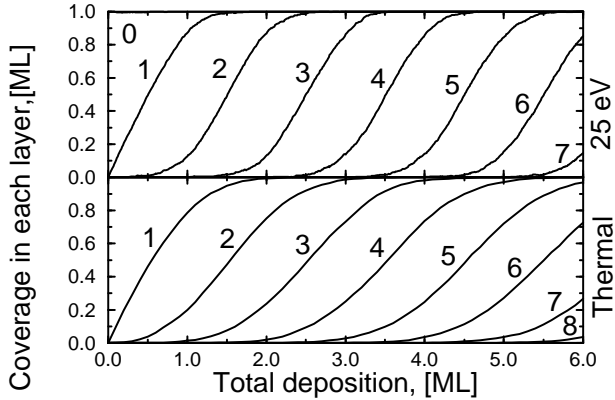}
    \caption{Plot of the coverage in each growing layer (layer \#
    labeled at each graph) above the
    initial surface as a function of the total deposition for
    Pt/Pt(111). Upper panel is for 25 eV, and lower panel is for thermal
    deposition.} 
    \label{fig:ptheightdist}
  \end{center}
\end{figure}

While the 18 eV and 25 eV energetic deposition has no influence on the 
surface morphology at low coverages, it has a significant influence 
as the growth progresses. Figure~\ref{fig:ptlayint1} shows fairly strong 
and slowly damped oscillations in the simulated anti--phase intensity
being induced by the energy. At 25 eV the periodicity in the growth 
of each layer
is evident from figure~\ref{fig:ptheightdist}
({\it i.e.,} the curves for successive layers look the same).
For thermal growth, on the other hand,
the completion of each layer spreads out over an increasing period of
time.

Two main factors play a role in making the energetic growth so smooth in the
case of Pt. First, figure~\ref{fig:ptstepimp} shows a pronounced
energy--induced mobility for impacts near steps that favors smooth growth,
{\it i.e.,} for impacts above the step, there is a high 
probability of energy--induced insertions. We do not see induced pile--ups 
for impacts below the step at these energies. Second, the step
density for Pt/Pt(111) at this temperature is very high. As was the case for
Ag we expect the advantageous effect of the energy to die away with increasing
surface temperature, as the step density decreases with increased surface 
diffusion.

\begin{figure}[htbp]
  \begin{center}
    \leavevmode
    \vspace*{3mm}
    \epsfxsize=8cm
    \vspace*{3mm}
% XXX Smaller
%    \epsffile{fig/fig23/5ml.ps}
    \epsffile{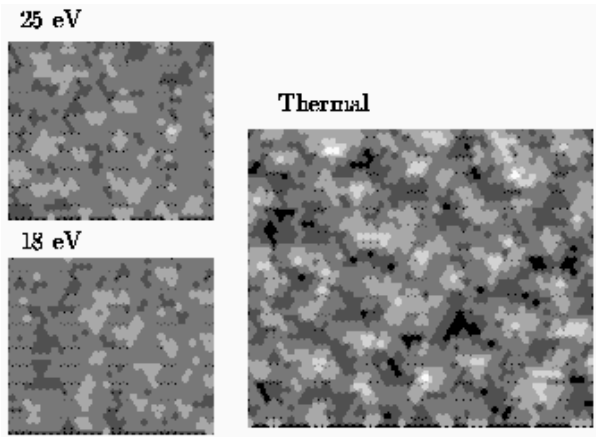}
    \caption{
Surface morphologies for Pt/Pt(111) at 7 meV (80 K) after a
    total deposition of 5 atomic layers, for thermal (size $50\times 50$)
    growth and energetic deposition at 18 (size $30\times 30$) and 25 eV
    (size $30\times 30$).}
    \label{fig:pt5ml}
  \end{center}
\end{figure}

\section{Discussion of results}
\label{sec:discres}

Here we summarize and discuss our results from simulating growth by energetic 
deposition. Our results from our study of the possible outcomes
of single impacts on different surface configurations
were summarized in section~\ref{sec:resultsmdsum}. We found energy--induced
defect formation mechanisms, such as adatom/vacancy formations,
pile--ups, step edge
restructurings and break--ups of existing islands.
We also found energy--induced defect annihilation mechanisms, in particular 
insertion of atoms into horizontally growing layers. For the energies
in this study we in general found the energy--induced mobility to be limited
to a range of a few ($\approx$5) atomic distances. 

In doing the KMC-MD simulations of the entire growth process, we have
seen that the energy--induced atomic mobility can affect island densities
in two ways. First, for energies above a certain threshold (20 eV for Ag), 
the energy--induced formation of adatom/vacancy pairs and direct formation
of dimers and trimers on the surface increases the island density. In our 
simulation these small clusters of atoms diffuse only slowly on the surface,
giving rise to the reduced island separation.
It is an important characteristic for islands formed in this manner that
they are nucleated by diffusion and aggregation: they tend to repel 
each other via their diffusional fields, and are well 
separated as is the case for islands formed in thermal 
growth. Hence, they coalesce in the late stages of the growth of a monolayer.
Second,
the energetic impacts increase the island density in a separate way,
by breaking off clusters of atoms from existing islands. While this
process may give rise to new islands, they are always nucleated very close
to existing islands, and are qualitatively different. In the Ag/Ag(111)
simulations, we have seen that they do not contribute to the step density,
in contrast to islands nucleated by diffusion.
They have a very short life time, either
because they diffuse to join the original island, or because they grow
and coalesce at a much earlier growth stage.

We have seen that at low temperatures using energetic deposition
can change rough thermal growth
to smooth layer--by--layer growth.
As usual, this is expected eventually to decay 
into rough growth as the number of layers increases~\cite{pulsesput}.
We have done simulations at temperatures where thermal interlayer mobility
is frozen out, and have found that energy--induced insertions of atoms into
growing layers
can be sufficient
to give layer--by--layer growth. However, energy--induced pile--ups,
which set in at a higher energy, can cause the opposite effect.
For this reason
we find that there is an optimal energy for layer--by--layer
growth --- for Ag it
is approximately 25 eV. We expect the existence of such an optimum
to be general.
Pile--ups are likely to dominate
over insertions at higher energies; for insertions
only the incoming atom or other atoms residing above descending steps
can be inserted, while the number of atoms that can pile up is not
similarly limited.

Since the energy--induced insertions are limited to impacts within a short
range of descending steps, we have seen that a high step density is 
needed for energy--induced smooth growth. While the use of energetic deposition
can itself assist in increasing the step density, we have seen the energy
induced layer--by--layer growth dies away with increasing temperature in the
case of Ag/Ag(111). The horizontal diffusivity for this system is very high,
and the two dimensional islands readily grow very large as the temperature
is increased, giving a very low step density. However, how
efficient the energetic beam is in increasing the island density
depends on many details of the atomic potential energy landscape.
In some cases this efficiency can be very high.
Therefore, in some systems which have a
low step density when thermally grown, it is possible
that an energetic beam might increase the step density sufficiently
for the energetic-beam insertions to give rise to smooth growth.

\section{Discussion of method}
\label{sec:discmeth}

We now turn to a general discussion of our Kinetic Monte--Carlo Molecular
Dynamics
method for simulating crystal growth by energetic deposition. We wish to address
the validity of the method, its advantages and disadvantages, and how we
find it useful in providing information about the energetic growth. 

Kinetic Monte--Carlo treats the thermal diffusion as a sequence of
uncorrelated atomic hops, and assumes these diffusion hops to be
instantaneous. In reality, a diffusion hop in a crystalline environment
has a duration time $\tau_m$ of roughly a picosecond, the typical time
an atom spends on top of the barrier while crossing it. The Kinetic
Monte--Carlo formalism is correct when the times between various kinds
of hops is much larger than $\tau_m$. (The time between hops is the inverse
rate of the diffusion process, given by the Arrhenius formula with the
barriers in Table~\ref{tab:modbar}.)
In the same way, KMC-MD is correct when the
time between hops is larger than $\tau_E$, the duration of our MD
simulation of the non-thermal collision--induced mobility. During
$\tau_E$ we neglect thermal mobility outside and across the boundaries
of the region of the MD simulation. This approximation only limits the
validity of the simulation to the extent that this mobility would
correlate to the atomic rearrangements in the impact region. The
collision is treated as instantaneous by the KMC algorithm, so the total
mobility outside the impact region is correctly accounted for, whereas
inside the region thermal mobility within $\tau_E$ is doubly accounted
for. While this should be kept in mind, we don't expect it to have any
significance for the simulations presented in this paper, because
diffusion is so slow compared to $\tau_E$.

The KMC--MD method also carries over the other usual limitations
of KMC. One of these is that the KMC part of the simulation must
be done on a lattice. In the present simulations of growth of fcc(111) 
surfaces, we neglect the presence of the off--lattice hcp binding 
sites in the KMC part of the simulation, and hence we have to move
atoms trapped in these sites at the end of an MD simulation to neighboring
fcc sites. This is severe only if growth in practice would take place
on the hcp sites. Growth partly on hcp sites, partly on fcc sites,
would give rise to complicated structures and dynamics at the stage
of coalescence
of these islands, a scenario we are unable to study with the KMC--MD method.
As noted earlier, the favorable hcp site is an artifact of the EMT potential
we use, and these issues are most likely not relevant experimentally.

The KMC--MD method must be based on a model potential for the atomic
interactions, in this case the EMT. While the EMT includes many
qualitative and to some extent quantitative features of the interaction 
of late transition and noble metals, it is not an exact potential.
In the case of Pt(111), for example, it is found that a simple scaling of 
the EMT energy barriers by a factor of 1.6 as input for a KMC simulation
of thermal growth gives a good agreement with experimental island densities,
transition of fractal to compact islands, and the appearance of reentrant
layer--by--layer growth at low temperatures~\cite{kmcpt2d3d,kmcptisl,ptisl}. 
A common scaling factor 
of all energy barriers would not be a severe discrepancy --- it merely
corresponds to a scaling of the temperature. 

While the approximations made in effective medium theory prevent us 
from controlled predictions of exactly how a real material would grow
by energetic deposition,
the method still includes the right ingredients to be a useful 
tool for our purposes: to examine how the use of energetic particles 
may influence the crystal growth; and to identify mechanisms and evaluate
their relative importance, as has been done in the previous sections.
We have demonstrated that the KMC--MD method can indeed be used to 
elucidate the important interplay between energy--induced mobility
and thermal surface diffusion.

One could imagine an alternative way of doing these simulations 
of growth by energetic depositions. First make
a complete table of all possible outcomes of single atom--surface
collisions listed together with their relative statistical significance,
and for all relevant local surface configurations. Then perform a KMC
simulation, which for each impact would choose the 
resulting local configuration from such a table. The present simulation 
study shows what an enormous task that would be. In section~\ref{sec:resultsmd} 
we considered impacts on the flat surface and near one type of straight step,
and
found a dependence on the step position up to 5 atomic distances away. 
Imagine the table of collision events needed to correctly account for the
evolution surfaces presented in section~\ref{sec:resultskmcmd}!

A major advantage of the present method is that it allows one to evolve the
surface
by depositing energetic atoms without making any assumptions about the
effect of the energy. From each simulation we can in principle record exactly
the contributions of various
types of energy--induced atomic mobilities. The method
correctly convolves the distribution of possible outcomes of energetic
impacts on a given local configuration with the distribution of
different local configurations during the crystal growth. The energy--induced
microscopic mechanisms acting are not assumed, but on the contrary revealed
by the method.  One example of this is the insertion of atoms residing atop
islands prior to impacts, as discussed in section~\ref{sec:atomatop}.

The main disadvantage of the method is that it is very time consuming
computationally. The MD simulations are expensive, and a new simulation
for every deposited atom is needed. Even for the relatively small 
surface areas considered here, a thousand to a few thousands of MD
simulations are needed per monolayer. In order for the method to be
feasible, it is essential that the island density not be too low.  Big
islands require a larger surface area and hence a larger number of MD 
simulations per monolayer. Our simulations ran for weeks to months on
reasonable 1997 workstations.

Our method is not well suited for parallelization. The bulk of the CPU time
is running the MD for the atomic impacts. 
Each MD simulation is small and parallelization
by subdivision would give rise to a considerable communication overhead.
The current size of the simulation is set by the need to have an impact
not be affected by the periodic boundary conditions: the impact and its
further diffusive evolution must not feel its periodic images.
Distribution of the impacts among different processors would necessarily
involve artificial simultaneous depositions in different areas of the surface.
For these depositions to be unaffected by one another, it would seem that
the same system-size per processor must be needed: parallelization does
not allow more monolayers per month.

Our new method, by combining MD for the impact with KMC for the diffusive
motion between impacts, allows for realistic simulations of surfaces grown
with energetic beams at realistic growth rates of monolayers per second.
Good simulations, however, take a week or more per monolayer deposited.

\section{Acknowledgements}
We thank K.W. Jacobsen, K. Bhattacharya, T. Curcic, and C. Henley for
helpful discussions. This research was supported by the Cornell Center
for Materials Research under NSF award number DMR-9632275.  The
simulations were done partially at the Cornell Theory Center, which
receives major funding from the National Science Foundation (NSF) and
New York State, with additional support from the National Center for
Research Resources at the National Institutes of Health (NIH), IBM
Corporation, and other members of the center's Corporate Partnership
Program.


\begin{references}
\bibitem{excuse} These simulations were done at 25 eV using MD setups with
  $\bar{N}=(3,5,2,4)$ and $\bar{\xi} = (0.03,0.06)$, which
  overestimates the adatom/vacancy production 
  somewhat. However, using converged MD system sizes, the same result
  would likely be obtained at a slightly higher energy, probably
  approximately 30-35 eV.
\end{references}
\end{document}